\documentstyle[11pt]{article}

\newcommand{\beq}{\begin{equation}}
\newcommand{\beqar}{\begin{eqnarray}}
\newcommand{\eeq}[1]{\label{#1} \end{equation}}
\newcommand{\eeqar}[1]{\label{#1} \end{eqnarray}}

\def\ess{\hskip.444444em plus .499997em minus .037036em}
\def\mss{\hskip.333333em plus .208331em minus .088889em}

\begin{document}    
\renewcommand{\topfraction}{0.95}
\renewcommand{\bottomfraction}{0.95}
\renewcommand{\textfraction}{0.05}
\pagestyle{myheadings}
\thispagestyle{empty}
\def\ti#1 {\begin{center} \baselineskip=17pt {\large #1} \end{center}}
\markboth           
{\it P. M\"{o}ller, J. R. Nix, and K.-L. Kratz/Nuclear
Properties}         
{\it P. M\"{o}ller, J. R. Nix, and K.-L. Kratz/Nuclear
Properties}         
\thispagestyle{empty}
\mbox{ } \hfill \mbox{ } \\
\mbox{ } \vspace{0.1in} \mbox{ } \\
\begin{center}      
\begin{large}       
{\bf NUCLEAR PROPERTIES FOR ASTROPHYSICAL APPLICATIONS}$^*$
\\[2ex]\end{large}  
P. M\"{O}LLER\\[1ex]
Theoretical Division, Los Alamos National
Laboratory\\        
Los Alamos, NM 87545\\
and\\               
Center for Mathematical Sciences, University of Aizu\\
Aizu-Wakamatsu, Fukushima 965-80, Japan\\[2ex]
J. R. NIX\\[1ex]    
Theoretical Division, Los Alamos National
Laboratory\\        
Los Alamos, NM 87545\\[2ex]
and\\[2ex]          
K.-L. KRATZ\\[1ex]  
Institut f{\"{u}}r Kernchemie, Universit{\"{a}}t Mainz\\
D-55099 Mainz, Germany \\[6ex]
January 1, 1996 \\[6ex]
\end{center}        
\newcounter{bean}   
\begin{list}        
{\Roman{bean}}{\usecounter{bean}
\setlength{\leftmargin}{0.5in}
\setlength{\rightmargin}{0.5in}
\setlength{\labelwidth}{0.25in}
\setlength{\labelsep}{0.25in}
}                   
\item[]             
We tabulate the ground-state odd-proton and
odd-neutron spins and parities, proton and neutron pairing gaps, binding energy,
one- and two-neutron separation energies,
quantities related to $\beta$-delayed
one- and two-neutron emission probabilities,
$\beta$-decay energy release and  half-life with
respect to Gamow-Teller decay,
one- and two-proton separation energies, and  $\alpha$-decay energy release and
half-life           
for 8979 nuclei     
ranging from $^{16}$O to $^{339}$136 and extending from the proton drip line
to the neutron drip line.
Single-particle level diagrams and other quantities are also presented in
graphical form.     
The starting point of our present work
is a study  of nuclear ground-state masses and deformations based
on the finite-range droplet model  and  folded-Yukawa single-particle
potential published in a previous issue of {\sc Atomic Data and Nuclear
Data Tables}.  The $\beta$-delayed neutron-emission probabilities
and Gamow-Teller $\beta$-decay rates are obtained from
a quasi-particle random-phase approximation with
single-particle levels and wave functions
at the calculated nuclear ground-state shapes as input quantities.
\\[1ex]             
\end{list}          
\vfill              
\begin{list}        
{\Roman{bean}}{\usecounter{bean}
\setlength{\leftmargin}{0.5in}
\setlength{\rightmargin}{0.5in}
\setlength{\labelwidth}{0.25in}
\setlength{\labelsep}{0.0in}
}                   
\item[]             
\begin{description} 
\item[\normalsize $^*$This paper is dedicated]\ to the memory of our friend and colleague Vilen M.\ Strutinsky,
who through his years of devoted research on nuclear-structure models,
most notably the Strutinsky shell-correction method, made possible many
of the calculations discussed here.
\end{description}   
\end{list}          
\newpage            
\tableofcontents    
\mbox{ } \vspace{-4.4pt} \mbox{ }\\
{{\bf EXPLANATION OF TABLE}}\  \hfill {\bf 141}\\[10.4pt]
{{\bf TABLE.}}\ess Calculated Nuclear Ground-State Properties \hfill {\bf 142}\\
\mbox{ } \vspace{-20.0pt} \mbox{ }\\
\section{INTRODUCTION}
 
In a previous issue of {\sc Atomic Data and Nuclear Data Tables}
we presented a calculation of nuclear ground-state masses and
deformations        
for 8979 nuclei     
ranging from $^{16}$O to $^{339}$136 and extending from the proton drip line
to the neutron drip line.$\,^{1})$\ess
The 1992 version of the finite-range droplet model and
folded-Yukawa single-particle potential that
was the basis for this calculation is referred to as the FRDM (1992).
We here use these ground-state
masses and deformations
as starting points  for calculations of
additional ground-state properties
that are useful for astrophysical and other applications.
 
An important feature of a mass model is its reliability for
nuclei beyond the region used for the determination
of the model constants. In particular, can one expect  the model
to be reliable for  nuclei very far from $\beta$-stability
and in the region of superheavy elements? In our mass
paper$\,^{1})$ we addressed the model reliability
for new regions of nuclei
by comparing predictions of masses that were
not included in the data set to which the model constants were
determined to new experimental data. Since
at that time we had available very few
new masses over and above the 1989 data set$\,^{2})$ from
which the model constants were determined, we had to test
model reliability by simulation. In one simulation, where the model
was adjusted to a 1977 experimental data set, it was found
that 351 new masses measured between 1977 and 1989 were calculated with
an increase of only 2\% in the model error. In another simulation,
where we adjusted the model constants only to nuclei
in the region $Z$, $N \geq 28 $ and
$A\leq 208$, we found that there was
no increase in the {\it mean} error but some increase in the
standard deviation for nuclei beyond $A=208$ that were not included in
the limited adjustment.
However, for superheavy nuclei such as
$^{288}$110 and $^{290}$110, the
difference in mass predictions between the FRDM (1992), whose
constants were determined by including nuclei up to $A=263$,
and the limited adjustment was only of the order of 1 MeV\@.
It is significant that these superheavy nuclei are 80 mass units
heavier than the heaviest nuclide included in the
limited adjustment. 
 
Because our previous studies indicated a very good model reliability
for new regions of nuclei, we here present calculations
of additional nuclear ground-state properties
based on the same model and the same values of model
constants,          
for the same set    
of 8979 nuclei considered in our mass calculation.$\,^{1})$\ess
Specifically, we consider the
following quantities:
\begin{description} 
\item[]             
{\bf Odd-nucleon spins and parities:}\\[2ex]
Projection of the odd-proton angular momentum along the
symmetry axis\\     
and parity of the wave function        \hfill $\Omega_{\rm p}^{\pi}$   \\[1ex]
Projection of the odd-neutron angular momentum along the
symmetry axis \\    
and parity of the wave function      \hfill $\Omega_{\rm n}^{\pi}$   \\
\item[]             
{\bf Lipkin-Nogami pairing gaps:}\\[2ex]
Proton pairing gap      \hfill $\Delta_{{\rm LN}_{\rm p}}$  \\[1ex]
Neutron pairing gap     \hfill $\Delta_{{\rm LN}_{\rm n}}$   \\
\item               
{\bf FRDM mass-related quantity:} \\[2ex]
Total binding energy             \hfill    $E_{\rm bind}$             \\
\item[]             
{\bf Neutron separation energies:} \\[2ex]
One-neutron separation energy              \hfill   $S_{\rm 1n}$ \\[1ex]
Two-neutron separation energy              \hfill   $S_{\rm 2n}$ \\
\item[]             
{\bf Beta-decay properties:}\\[2ex]
Probability for producing a final nucleus with mass number $A$\\
following $\beta$ decay and delayed neutron emission
                                                       \hfill $P_{A}$ \\[1ex]
Probability for producing a final nucleus with mass number $A-1$\\
following $\beta$ decay and delayed neutron emission
                                                       \hfill $P_{A-1}$ \\[1ex]
Probability for producing a final nucleus with mass number $A-2$\\
following $\beta$ decay and delayed neutron emission
                                                       \hfill $P_{A-2}$ \\[1ex]
Energy released in $\beta$ decay        \hfill $Q_{\beta}$  \\[1ex]
Half-life  with respect to Gamow-Teller
$\beta$ decay               \hfill $T_{\beta}$\\
\newpage            
\item[]             
{\bf Proton separation energies:} \\[2ex]
One-proton separation energy               \hfill   $S_{\rm 1p}$ \\[1ex]
Two-proton separation energy               \hfill   $S_{\rm 2p}$ \\
\item[]             
{\bf Alpha-decay properties:}\\[2ex]
Energy released in  $\alpha$ decay       \hfill $Q_{\rm \alpha}$   \\[1ex]
Half-life  with respect to $\alpha$ decay         \hfill $T_{\alpha}$ \\
\end{description}   
 
The details of the calculations are given in Sec.~2.
Separation energies and energy releases are readily obtained from
mass differences. The $\beta$-decay half-lives and
$\beta$-delayed neutron-emission probabilities are obtained
from a quasi-particle random-phase approximation
(QRPA). In the QRPA 
the single-particle energies and wave functions at
the calculated ground-state deformation serve as the starting point.
The tabulated results are described in Sec.\ \ref{sec3}.
 
After the           
submission of our mass paper$\,^{1})$ a new mass
evaluation$\,^{3,4})$ has become available. It contains
217 new masses that were not included in the 1989 data set$\,^{2})$
from which the values of the FRDM (1992) constants were determined.
Therefore, we are now able to assess the reliability, without
simulations, of the FRDM (1992) and of several other
models that are also commonly used in astrophysical calculations.
These reliability issues are discussed in Sec.~\ref{reli}.
 
As one particular application, we discuss in Sec.~\ref{astro} the use
of the calculated quantities in astrophysical {\it rp\/}- and
$r$-process calculations.  However,  our results also have wider
applicability to several other areas of astrophysics and to such fields
as reactor physics. Section~\ref{singlev} contains calculated proton
and neutron single-particle level diagrams for representative spherical
and deformed nuclei throughout the periodic system.
 
\section{CALCULATIONAL DETAILS}
 
The quantities studied in this paper are obtained in
four different ways.
\begin{enumerate}   
 
\item               
The odd-proton spin and parity $\Omega_{\rm p}^{\pi}$,
odd-neutron spin and parity
${\Omega_{\rm n}^{\pi}}$, proton pairing gap
$\Delta_{{\rm LN}_{\rm p}}$, and
neutron pairing gap 
$\Delta_{{\rm LN}_{\rm n}}$
are microscopic quantities obtained simultaneously with the
calculated ground-state masses and deformations.
They were not published in our mass paper
because of space limitations.
 
\item               
The total binding energy $E_{\rm bind}$, one-neutron separation energy
$S_{\rm 1n}$, two-neutron separation energy $S_{\rm 2n}$,
$\beta$-decay energy release $Q_{\beta}$, one-proton
separation energy $S_{\rm 1p}$, two-proton separation energy $S_{\rm 2p}$,
and  $\alpha$-decay energy release $Q_{\alpha}$ are  obtained
from appropriate differences of
the calculated mass excesses. For convenient
access we publish them here.
 
\item               
The $\beta$-delayed occupation
probabilities $P_{A}$, $P_{A-1}$, and $P_{A-2}$
and $\beta$-decay half-lives $T_{\beta}$ are obtained
from a microscopic quasi-particle random-phase approximation (QRPA).
 
\item               
The $\alpha$-decay half-life $T_{\alpha}$ is obtained from
the semi-empirical relationship of Viola and Seaborg,$\,^{5})$\mss
with constants determined by Sobiczewski, Patyk, and
{\v{C}}wiok.$\,^{6})$\ess
\end{enumerate}     
 
\subsection{Odd-nucleon spin and parity}
 
The odd-nucleon spin is simply the
projection of the angular momentum along the symmetry axis
($\Omega$ quantum number)
for the last occupied proton or neutron
level, when this level is occupied by a single nucleon.
For odd-proton or   
odd-neutron nuclei the spin of the nucleus is simply $\Omega_{\rm p}$
or $\Omega_{\rm n}$, respectively. The superscript $\pi$ gives
the parity of the wave function.
 
For spherical nuclei with degenerate levels, the nuclear spin
is defined as the maximum value of $j_z$, which is $|j|$.
Thus, for a spherical nucleus we cannot use as an odd-even
spin assignment the $\Omega$ value automatically provided for
the last occupied single-particle level, since this level
is randomly assigned any $\Omega$ value in the range
1/2 to $|j|$. For slightly deformed nuclei one
could in principle use deformed assignments, but in practice
it would be unrealistic to list a deformed assignment for
a nucleus with a calculated deformation of, say,
$\epsilon_2=0.01$. We therefore proceed in the following manner.
For nuclei with a deformation $|\epsilon_2|\geq\epsilon_{\rm crit}$
a deformed assignment is used. For nuclei with
$|\epsilon_2| < \epsilon_{\rm crit}$ we calculate the levels
for a spherical shape and adopt the spherical spin assignment thus obtained.
When we compare below in Sec.~\ref{gspin} calculated odd-particle
spins and parities with experimental data, we show that the results are quite
insensitive to the exact value of $\epsilon_{\rm crit}$. We choose here
\begin{center}      
\begin{tabular}{rcl}
$\epsilon_{\rm crit}$ & = & 0.15 \\[2ex]
\end{tabular}       
\end{center}        
For nuclei whose ground states are calculated to have the octupole
shape parameter $\epsilon_3\neq 0$ parity is not conserved and is
therefore not tabulated.

\subsection{Pairing gaps}
 
In an extensive study of nuclear pairing$\,^{7})$
we investigated both a macroscopic
pairing model and a microscopic pairing model, which was solved in both
the BCS$\,^{8-11})$ and
Lipkin-Nogami  (LN)$\,^{12-14})$ approximations.
For each model we determined a preferred form of the effective pairing
interaction and optimum values of the constants of the effective pairing
interaction, which were obtained from a least-squares minimization of the
difference between calculated pairing gaps and experimental odd-even
mass differences.   
 
An important result of our previous study is that it is crucial to
differentiate between several pairing-gap concepts. The most simple
concept is the {\it average} pairing gap $\overline{\Delta}$, which is an
algebraic relationship such as $c/\sqrt{A}$, where $c$ is a constant
and $A$ is the number of nucleons in the system being studied.
The average pairing 
gap may be regarded as a macroscopic model for the nuclear
pairing gap, and it may therefore be directly compared with
experimental odd-even mass differences.
 
When a microscopic approach is used the situation is considerably more
complicated. In this case the quantities that are compared to
experimental odd-even mass differences are obtained as  solutions to
microscopic pairing equations, for example the BCS or LN equations.
In the BCS method   
it is $\Delta$ that should be directly compared to the odd-even
mass differences. However,
in the LN approximation it is the sum of the pairing gap $\Delta$ and
the number-fluctuation constant $\lambda_2$,
where $\Delta$ and $\lambda_2$ are obtained as
solutions of the LN equations, that should be compared
to odd-even mass differences. We denote this sum by $\Delta_{\rm LN}$.
Thus,  $\Delta_{\rm LN}= \Delta + \lambda_2$.
 
To solve the usual pairing equations$\,^{7})$
one needs in addition to
single-particle energies also the value of the pairing-strength
constant $G$. This constant depends in a complicated way on the
number of levels included in the calculation and on the particular
nuclear region considered. However, it may be determined from an
effective-interaction pairing gap $\Delta_G$ by use of a
Strutinsky-like procedure.$\,^{7})$\ess
At first sight this may seem  an
unnecessary complication, but the advantage is that $\Delta_G$ does not
depend on the particular truncation of the single-particle level
spectrum that is chosen in the calculation. Furthermore, it depends in
a very simple way on $Z$ and $N$. Therefore, a
significant simplification is achieved if one considers $\Delta_G$ to
be the primary input quantity for pairing calculations,
with the            
constants that enter the function that defines $\Delta_G$  to be
the pairing-model effective-interaction constants.
 
In our earlier study$\,^{7})$ we obtained  the following preferred
functional form for the effective-interaction pairing gap $\Delta_G$:
 
\beqar              
{\Delta_G}_{\rm p} = \frac{rB_{\rm s}}{Z^{1/3}}e^{-tI^2} \nonumber \\[1ex]
{\Delta_G}_{\rm n} = \frac{rB_{\rm s}}{N^{1/3}}e^{-tI^2}
\eeqar{finalgap}    
Here $Z$ and $N$ are the numbers of protons and neutrons, respectively,
$I=(N-Z)/(N+Z)$ is the relative neutron excess,
and $B_{\rm s}$ is the surface area of the nucleus divided by
the surface area of the spherical shape.
From root-mean-square minimizations
we obtained results 
consistent with $t=0$  for both the BCS and LN models.
For these cases Eq.~(\ref{finalgap}) simplifies to
 
\beqar              
{\Delta_G}_{\rm p} = \frac{rB_{\rm s}}{Z^{1/3}} \nonumber \\[1ex]
{\Delta_G}_{\rm n} = \frac{rB_{\rm s}}{N^{1/3}}
\eeqar{finalgap2}   
and the effective-interaction pairing gap $\Delta_{G}$ is determined by
one constant for the entire nuclear chart, for both
protons and neutrons.
 
In our nuclear mass calculation$\,^{1})$
we performed a refined determination of the effective
pairing-interaction constant $r$, with the result that we adopted
the value $r=3.2$~MeV instead of the earlier value$\,^{7})$
$r=3.3$ MeV\@.  For details we refer to these earlier
studies.$\,^{1,7})$\ess
Although we revised the effective-interaction
pairing constant $r$ by 3\%, the earlier extensive pairing
study$\,^{7})$ can still serve as an excellent guide to the
properties of our current pairing model. Below we present further results
obtained in the current model.
 
\subsection{Total binding energy}
 
The total binding energy $E_{\rm bind}(Z,N)$ is
related to the atomic mass excess $M(Z,N)$ through the
simple relationship 
\beq                
E_{\rm bind} = ZM_{\rm H} + NM_{\rm n} - M(Z,N)
\eeq{bind}          
where               
$M_{\rm H}$ is the hydrogen-atom mass excess
and                 
$M_{\rm n}$ is the neutron mass excess. The total binding energy
includes the binding energy of the $Z$ electrons comprising the
atom, which we approximate by $a_{\rm el}Z^{2.39}$, with
$a_{\rm el} = 1.433\times 10^{-5}$~MeV\@.
 
For the benefit of workers in other fields, who are often confused by the
conventions adopted in atomic masses, we mention that the reason that the
atomic mass excess is tabulated instead of the atomic mass itself is simply to
eliminate the repetitive tabulation of additional leading numbers that can
easily be restored by adding the mass number $A$ times the mass unit u, which
is 1/12 the mass of the $^{12}$C atom, to the tabulated quantity.  Also, the
reason that the atomic mass is considered rather than the nuclear mass is that
the former is the actual experimentally measured quantity, whereas the latter
is less accurate because its extraction requires a knowledge of the binding
energy of the $Z$ atomic electrons.
 
For those applications where it is necessary to know the actual mass of the
nucleus itself, its value (in MeV) can be found from the atomic mass excess
tabulated in Ref.$\,^{1})$ by use of the relationship
\beq                
M_{\rm nucleus} = A {\rm u} + M - Zm_{\rm e} + a_{\rm el}Z^{2.39}
\eeq{nuclmass}      
where $m_{\rm e} = 0.51099906$~MeV is the mass of the
electron.$\,^{15,16})$\ess
As discussed in Ref.$\,^{1})$, the value ${\rm u} =
931.5014$~MeV that was used in the interim 1989 mass evaluation$\,^{2})$ should
be used for the atomic mass unit in Eq.~(\ref{nuclmass}).
 
\subsection{Neutron separation energies}
 
The one- and two-neutron separation energies $S_{\rm 1n}(Z,N)$ and
$S_{\rm 2n}(Z,N)$ are obtained from the mass excesses
through the differences
\beqar              
S_{\rm 1n}(Z,N) & = &  M(Z,N-1) +  M_{\rm n} - M(Z,N) \nonumber \\
S_{\rm 2n}(Z,N) & = &  M(Z,N-2) + 2M_{\rm n} - M(Z,N)
\eeqar{nsepar}      
 
\subsection{$\beta$-decay properties}
 
The formalism we use to calculate Gamow-Teller (GT) $\beta$-strength
functions is fairly lengthy, since it involves
adding  pairing and Gamow-Teller residual interactions
to the folded-Yukawa single-particle Hamiltonian and
solving the resulting Schr\"{o}dinger equation in the
quasi-particle random-phase approximation. Because
this model has been completely described
in two previous papers,$\,^{17,18})$\mss
we refer to those two publications for a full
model specification and for a definition of notation used.
We restrict the discussion here to an overview of features that
are particularly relevant to the results discussed in
this paper.

It is well known that wave functions and transition
matrix elements are more affected
by small perturbations to the Hamiltonian than are the
eigenvalues. When transition rates are
calculated it is therefore necessary to add
residual interactions to the folded-Yukawa single-particle
Hamiltonian in addition to the pairing interaction that
is included in the mass model. Fortunately,
the residual interaction may be restricted
to a term specific to the particular
type of decay considered. To obtain reasonably accurate half-lives
it is also very important to include ground-state deformations. Originally the
QRPA formalism was developed for and applied only to spherical
nuclei.$\,^{19,20})$\ess The extension to
deformed nuclei, which is necessary in global calculations of
$\beta$-decay properties, was first described in 1984.$\,^{17})$\ess
 
To treat Gamow-Teller $\beta$ decay  we therefore add
the Gamow-Teller  force
\beq                
V_{\rm GT}=2\chi_{\rm GT}:\mbox{\boldmath $\beta^{1-}\cdot\beta^{1+}$}\!:
\eeq{vgt}           
to the folded-Yukawa single-particle Hamiltonian,
after pairing has already been
incorporated,       
with the standard choice $\chi_{\rm GT}=
23$~MeV/$A$.$\,^{17-20})$\ess
Here \mbox{\boldmath $\beta^{1\pm}$}$= \sum_i$\mbox{\boldmath
$\sigma_it_i^{\pm}$} are the Gamow-Teller $\beta^{\pm}$-transition operators,
and the colans mean that all contractions in the quasi-particle
representations of the enclosed operator are to be ignored.
The correlations generated by the GT force are of specific
importance to the Gamow-Teller decays, which are the dominant decay
modes in many nuclei of astrophysical interest. Other types of residual
interactions are of importance for other decay modes, but leave
the Gamow-Teller decay rates unaffected, and can consequently be ignored
for our present purpose.
 
It should be noted that the RPA treatment formulated by Halbleib and
Sorensen$\,^{20})$ incorporates only particle-hole correlations
of specific importance to GT transitions. It has been
proposed$\,^{21,22})$ that the effect of neglected
particle-particle terms may be significant for $\beta^+$ transitions.
We later address this question in Sec.~\ref{betcal}.
Moreover, the RPA treatment may not contain enough ground-state
correlations.$\,^{22})$\ess However, in view of the present
uncertainties regarding these points we leave possible further refinements
for future consideration. Some additional comments are made
in Sec.~\ref{betcal}.

We next discuss the calculation of $\beta$-decay half-lives for
Gamow-Teller decay and the related problem of calculating
$\beta$-delayed neutron-emission probabilities.
In our discussion of the model we use, unless otherwise stated, expressions
and notation from   
the books by deShalit and Feshbach$\,^{23})$ and
Preston$\,^{24})$ and from our previous
publications.$\,^{17,18})$\ess
 
\subsubsection{$\beta^{-}$ and $\beta^{+}$ decay}
 
The process of $\beta$ decay occurs from an initial ground state or excited
state in            
a mother nucleus to a final state in the daughter nucleus.
For $\beta^-$ decay, the final configuration
is a nucleus in some excited state or its ground state,
an electron (with energy $E_{\rm e}$), and an antineutrino (with energy
$E_{\nu}$).         
The transition from the initial to the final state then involves
an operator $H$, which is the weak-interaction Hamiltonian density.
Once the operator $H$ is known, the probability per unit time
for emitting an electron with momentum
between $\hbar \mbox{\boldmath $k$}_{\rm e}$ and
$\hbar(\mbox{\boldmath $k$}_{\rm e} +  d\mbox{\boldmath $k$}_{\rm e})$
and an antineutrino with  momentum between
$\hbar \mbox{\boldmath $k$}_{\nu}$ and
$\hbar(\mbox{\boldmath $k$}_{\nu} +  d\mbox{\boldmath $k$}_{\nu})$
is given by the well-known Golden Rule
\beq                
dw_{fi} = \frac{2\pi}{\hbar}|H_{fi}|^2
\frac{ d\mbox{\boldmath $k$}_{e  }}{(2\pi)^3}
\frac{ d\mbox{\boldmath $k$}_{\nu}}{(2\pi)^3}\,
\delta(E_0 - E_{\rm e} - E_{\nu})
\eeq{golden}        
where $E_0$ is the energy released in the decay.
 
In the above expression one should sum
over the spins of the final states and average
over the initial spins. Our interest here is mainly to
obtain the probability of decay to a specific final nuclear state
$f$. To obtain this probability
one must go through several lengthy
steps. These steps are usually glossed over in discussions
of these models, but one fairly extensive account of
these steps is given in the book
by Preston.$\,^{24})$\ess
The final expression obtained through these steps
for the total probability for decay to one nuclear state is
\beq                
w_{fi} = \frac{m_0c^2}{\hbar} \;
\frac{\Gamma^2}{2\pi^3}
\;|M_{fi}|^2        
f(Z,R,\epsilon_0)   
\eeq{distint}       
where               
$\epsilon_0=E_0/m_0c^2$, with
$m_0$  the electron mass.
For consistency with standard treatments of $\beta$ decay
we here use SI units.
Moreover, $|M_{fi}|^2$ is the nuclear matrix element,
which is also the $\beta$-strength function.
The dimensionless constant $\Gamma$ is defined by
\beq                
\Gamma \equiv \frac{g}{m_0c^2} {\left( \frac{m_0c}{\hbar} \right)}^3
\eeq{gamma}         
where $g$ is the Gamow-Teller coupling constant.
There is a misprint  concerning this quantity in
the book$\,^{23})$ by deShalit and Feshbach, where in
their Chapter~9,    
Eq.~(2.11) the exponent is erroneously given as 2 instead of
the correct value 3.
The quantity $f(Z,R,\epsilon_0)$ has been extensively discussed
and tabulated elsewhere.$\,^{23-25})$\ess
 
For the special case in which
the two-neutron separation energy $S_{\rm 2n}$ in the daughter nucleus
is greater          
than the energy $Q_{\beta}$  released in the decay,
the probability for $\beta$-delayed one-neutron emission,
in percent, is given by
\beq                
P_{\rm 1n} = 100 \, \frac
{\begin{displaystyle}\sum_{S_{\rm 1n}<E_f<Q_{\beta}} w_{fi}\end{displaystyle}}
{\begin{displaystyle}\sum_{0<E_f<Q_{\beta}} w_{fi}\end{displaystyle}}
\eeq{deln}          
where               
$E_{f}$ is the excitation energy in the daughter nucleus
and $S_{\rm 1n}$ is the one-neutron
separation energy in the daughter nucleus.
We assume that decays to energies above
$S_{\rm 1n}$ always lead to delayed neutron emission.
In the more general case where multiple-neutron
emission is energetically possible,
but under the restriction that $S_{\nu{\rm n}}$
in the daughter nucleus
monotonically increase with increasing $\nu$, we define
\beq                
P_{\nu{\rm n}}^> = 100 \, \frac
{\begin{displaystyle}\sum_{S_{\nu{\rm n}}<E_f<Q_{\beta}} w_{fi}\end{displaystyle}}
{\begin{displaystyle}\sum_{0<E_f<Q_{\beta}} w_{fi}\end{displaystyle}}
\eeq{deln2}         
where $P_{\nu{\rm n}}^>$ is the probability of emitting $\nu$ or more neutrons.
The occupation probabilities $P_{A-\nu}$ introduced above are then
given by            
\beqar              
P_{A} &= &100 - P_{\rm 1n}^>  \nonumber \\[1ex]
P_{A-\nu} &= &P_{\nu{\rm n}}^> -  P_{(\nu +1) {\rm n}}^>
\eeqar{deln4}       
 
For some very neutron-rich nuclei
$S_{\nu{\rm n}}$  no longer
monotonically increase with increasing $\nu$,
in which case the above formalism
cannot be used. In the Table we present calculated values  only when
the above formalism is valid and  otherwise put ``\dots'' in the
columns for $P_{A-\nu}$.

To obtain the half-life with respect to $\beta$-decay
one sums up the decay rates $w_{fi}$ to the individual nuclear
states in the allowed energy window. The half-life is then related to the total
decay rate by       
\beq                
T_{\beta} = \frac{\ln 2}{\begin{displaystyle}\sum_{0<E_f<Q_{\beta}}
 w_{fi}\end{displaystyle}}
\eeq{beha}          
The above equation may be rewritten as
\beq                
T_{\beta}  =  \frac{\hbar}{m_0c^2} \; \frac{2\pi^3\ln 2}{\Gamma^2}
\frac{1}{\begin{displaystyle}\sum_{0<E_f<Q_{\beta}} \;|M_{fi}|^2
f(Z,R,\epsilon_0)   
\end{displaystyle}} 
 =  \frac{B}{\begin{displaystyle}\sum_{0<E_f<Q_{\beta}} \;|M_{fi}|^2
f(Z,R,\epsilon_0)   
\end{displaystyle}} 
\eeq{beha2}         
with                
\beq                
B  = \frac{\hbar}{m_0c^2} \; \frac{2\pi^3\ln 2}{\Gamma^2}
\eeq                
{beha3}             
For the value of $B$ corresponding to Gamow-Teller decay
we use              
\beq                
B= 4131 \; {\rm s}  
\eeq{bvalue}        
 
The energy released in electron
emission is         
\beq                
Q_{\beta^-}=E_0^{\beta^-} = \left[ M(Z,N) - M(Z+1,N-1)\right]c^2
\eeq{erelel}        
whereas the energy released in positron
emission is         
\beq                
Q_{\beta^+}=E_0^{\beta^+} = \left[ M(Z,N) - M(Z-1,N+1) -2m_0 \right]c^2
\eeq{erelpos}       
 
The above formulas apply to $\beta^-$ and  $\beta^+$ decay.
However, for  calculating half-lives
electron capture (EC) must also be considered.
 
\subsubsection{Electron capture}
 
The energy released in
electron capture is 
\beq                
Q_{\rm EC}=E_0^{\rm EC} = \left[ M(Z,N) - M(Z-1,N+1) \right] c^2
- {\rm electron\; binding\;  energy}
\eeq{erelec}        
so that             
\beq                
E_0^{\rm EC} = E_0^{\beta^+} + 2m_0c^2 - {\rm electron\;  binding\;  energy}
\eeq{difecpo}       
This shows that for some decays electron capture is possible
whereas $\beta^+$ decay is energetically forbidden.
The total probability for decay to one nuclear state is
again given by Eq.~(\ref{distint}), where the final state $f$ now refers
to electron capture over all electron shells {\it or} to $\beta^+$ decay.
 
The total half-life with respect to $\beta^+$ and EC decay
is given by         
\beq                
T_{\beta} = \frac{\ln 2}{\begin{displaystyle}
\left(\sum_{0<E_f<Q_{\rm EC}} w_{fi}^{\rm EC} + \sum_{0<E_f<Q_{\beta^{+}}}
 w_{fi}^{\beta^+}\right) \end{displaystyle}}
\eeq{thecpo}        
As pointed out above, the energies involved
in the two terms in the sum differ by $2m_0c^2$ minus the electron
binding energy      
and for some nuclear
final states  $w_{fi}^{\beta^+}$ may be zero (energetically forbidden)
while $w_{fi}^{\rm EC}$ is not.
 
To obtain an initial feel for these models we first studied the
approximate relativistic expressions given by
Preston.$\,^{24})$\ess The results obtained by use of these
expressions for $\beta^-$ decay are typically within 20\% of those
obtained in the more exact treatment by Gove and
Martin,$\,^{25})$\mss who have made extensive tabulations of
$f(Z,R,\epsilon_0)$.  Despite the small differences we have obtained
the computer code used to generate the tables of Gove and Martin and
incorporated it into our programs.  The results presented here have
been obtained with this more accurate treatment.
 
In the Table we present five quantities related to $\beta$ decay,
namely $P_{A}$, $P_{A-1}$, $P_{A-2}$, $Q_{\beta}$, and $T_{\beta}$. The
precise meaning of these quantities is as follows. When both $\beta^+$
and $\beta^-$ decay are possible, we tabulate $\pm$ in the columns for
$Q_{\beta}$ and $T_{\beta}$.  When neither $\beta^+$ nor
$\beta^-$ decay is possible, we tabulate ``\dots'' in the column for
$Q_{\beta}$, ``$\beta$-stable'' in the column for  $T_{\beta}$, and
blank fields in the columns for $P_{A}$, $P_{A-1}$, and $P_{A-2}$.
When only EC or $\beta^+$ decay is possible, we tabulate $Q_{\rm EC}$
in the column for $Q_{\beta}$, the calculated half-life with respect to
Gamow-Teller decay for combined EC and $\beta^+$ decay in the column for
$T_{\beta}$, and blank fields in the columns for $P_{A}$, $P_{A-1}$,
and $P_{A-2}$.  Finally, when only  $\beta^-$ decay is possible, we
tabulate $Q_{\beta^{-}}$ in the column for $Q_{\beta}$, the calculated
half-life with respect to Gamow-Teller $\beta^-$ decay in the column
for  $T_{\beta}$, and the calculated occupation probabilities after
$\beta$-delayed neutron emission in the columns for $P_{A}$, $P_{A-1}$,
and $P_{A-2}$. The electron binding energy has been neglected in
the determination of $Q_{\rm EC}$.
 
To obtain more accurate values of $T_{\beta}$ and
$P_{\nu {\rm n}}$, we have
calculated the $Q$ values that enter Eqs.~(\ref{deln2}), (\ref{beha}), and
(\ref{thecpo}) from experimental mass differences when all experimental masses
that are required for  the calculation are available and otherwise
from calculated mass differences. The neutron separation energies that enter
in the calculations of $P_{\nu {\rm n}}$  in Eq.~(\ref{deln2})
are also obtained from experimental mass differences when available and
otherwise from calculated mass differences. Calculated deformations are always
used, even when experimental data are available. However, in the astrophysical
applications presented below, further use is made of experimental information,
as is discussed in Sec.~\ref{astro}.

\subsection{Proton separation energies}
 
The one- and two-proton separation energies $S_{\rm 1p}(Z,N)$ and
$S_{\rm 2p}(Z,N)$ are obtained from mass excesses
through the differences
\beqar              
S_{\rm 1p}(Z,N) & = &  M(Z-1,N) +  M_{\rm H}  - M(Z,N)\nonumber \\
S_{\rm 2p}(Z,N) & = &  M(Z-2,N) + 2M_{\rm H}  - M(Z,N)
\eeqar{psepar}      
 
\subsection{$\alpha$-decay properties}
 
The five heaviest known elements 107, 108, $_{109}$Mt, 110, and 111
were all identified from their $\alpha$-decay
chains,$\,^{26-30})$\mss
which limited their stability. The neutron number of the first
identified isotope of 108 and Mt was $N=157$.  Because $\alpha$-decay
chains provide very clear signatures of the nuclear species in the
beginning of the decay chain and fission does not, it is likely that
additional new nuclei discovered in the heaviest region will often be
identified through their $\alpha$-decay properties.  Models of
$\alpha$-decay properties are therefore highly useful for designing and
interpreting experiments that explore the limits of stability of the
heaviest elements.  Obviously, one also needs to consider whether
spontaneous-fission half-lives are significantly shorter than the
$\alpha$-decay half-lives.  In that case spontaneous-fission would be
the dominating decay mode and $\alpha$ decay might not be detected.
 
The single most important quantity determining the
$\alpha$-decay      
half-life is the energy release $Q_{\alpha}$.\ess  In the heavy-element region
an uncertainty of 1 MeV in  $Q_{\alpha}$  corresponds to an uncertainty
of $10^{\pm 5}$ for $Q_{\alpha}\approx 7$ MeV and to an uncertainty of $10^{\pm 3}$ for
$Q_{\alpha}\approx 9$ MeV\@.$\,^{31})$\ess
The energy release $Q_{\alpha}$ is obtained from the mass excesses through
the difference      
\beq                
Q_{\alpha}(Z,N) = M(Z,N) - M(Z-2,N-2) - M(2,2)
\eeq{qalph}         
 
The $\alpha$-decay half-lives $T_{\alpha}$
presented in the Table are estimated by
use of the Viola-Seaborg relationship$\,^{5})$
\beq                
\log (T_{\alpha}/{\rm s}) = (aZ+b){(Q_{\alpha}/{\rm MeV})}^{-1/2}+(cZ+d)
\eeq{violas}        
where $Z$ is the proton number of the parent nucleus.  Instead of using
the original set of constants suggested by Viola and Seaborg we use
the more recent values
\beqar              
\begin{array}{ll}   
a=\mbox{}+1.66175, & b=\phantom{0}\mbox{}-8.5166 \nonumber\\
c= \mbox{}-0.20228, & d=\mbox{}-33.9069
\end{array}         
\eeqar{parviol}     
that were determined in an adjustment taking into account new data for
even-even nuclei.$\,^{6})$\ess The uncertainties in the
calculated half-lives due to this semi-empirical approach are far smaller
than uncertainties due to errors in the calculated energy release.

\section{TABULATED RESULTS \label{sec3}}
 
Deformed single-particle models provide the  starting point
 for the calculations of nuclear ground-state masses
and deformations, which were extensively discussed
in our previous paper.$\,^{1})$\ess
Since               
nuclear wave functions are also provided by these models,
one may also use these models to determine
electromagnetic moments and
transition rates,   
$\beta$-strength functions, $\beta$-decay half-lives,
and $\beta$-delayed neutron-emission probabilities.
 
The results of our calculations of many such
nuclear properties  for astrophysical applications are
presented in the Table. To provide an
overview of these  results,
we present in Figs.~\ref{lippdpl2}--\ref{rproc} color
diagrams of the calculated
pairing gaps, neutron separation energies,
proton separation   
energies, $\alpha$-decay energy release and half-life,
$\beta$-decay half-life,
and some related quantities. Beyond $Z\approx 120$ and $N\approx 190$
the calculated potential-energy surfaces on which
Figs.~\ref{lippdpl2}--\ref{rproc} are based are very flat and the barrier with
respect to fission is almost zero.
Because the  ground state is identified
as the deepest local pocket in this flat surface, one obtains
rapidly fluctuating deformations, separation energies, and
energy releases between
neighboring nuclei. These results are of no physical significance, since
the spontaneous-fission half-life of these nuclei will be much too short
to allow experimental observation.

\subsection{Odd-nucleon spin and parity\label{gspin}}
 
The most important constants in the folded-Yukawa single-particle
model are the diffuseness and spin-orbit constants, which were
determined$\,^{32})$ in 1974 in the rare-earth and actinide
regions from comparisons between calculated and experimental
single-particle level orderings.  The global nuclear-mass
study$\,^{33})$ in 1981 introduced a  set of constants valid for
the entire nuclear chart in terms of an expression for the spin-orbit
strength that is linear in $A=N+Z$, with the expression fully defined
by the previously determined  values in the actinide and
rare-earth regions. This  procedure is somewhat
subjective because it is not based on exact comparisons between all
available experimental data and calculations.
Instead, one typically proceeds by
calculating single-particle level diagrams as functions of deformation
for several sets of constants, comparing their structure to a few
selected nuclei, and choosing the set
that gives the best agreement.
 
Because we now have available nuclear ground-state shapes from our
calculations of ground-state masses, we are in a position to compare
calculated and experimental ground-state spins and parities in a well-defined
manner, as shown in Figs.~\ref{spincoa92a} and \ref{spincob92a}.  The
only ambiguity concerns the  comparison for nuclei calculated to be
weakly deformed. We have chosen to base the comparison on spherical
assignments if $|\epsilon_2|<0.15$ in the calculations. With this
choice we obtain agreement in 446 cases and disagreement in 267 cases,
corresponding to 63\% agreement.  When the ground-state energy is {\it
not\/} minimized with respect to $\epsilon_3$ and $\epsilon_6$,  we
obtained in a previous calculation, as expected, slightly less
favorable results:  agreement in 428 cases and disagreement in 285
cases, corresponding to 60\% agreement.  These results are not very
sensitive to changes in the choice concerning when to use spherical
assignments. In the present study, if we {\it always} choose the spherical
assignment when  this choice yields agreement with experimental data,
we obtain agreement in  482 cases and disagreement in 231 cases,
corresponding to 68\% agreement,
so that the improvement in the agreement is only 5\%.
 
The                 
disagreements between the calculated and experimental spins and parities usually
arise because several deformed or spherical levels lie very close
together, making accurate calculations difficult.  For magic numbers
there is an almost stunning agreement, which, taken together with our
analysis of the disagreements in other regions, makes it unlikely that
a significantly better {\it global} set of constants can be found.  The
existing disagreements probably arise from
residual interactions outside the framework of the single-particle
model.              
 
\subsection{Pairing gaps}

We  first emphasize the most important results of our previous
pairing             
study:$\,^{7})$     
\begin{enumerate}   
 
\item               
The preferred form of the pairing gap given by
Eq.~(\ref{finalgap2}) lowers the rms deviation by
about 20\% relative to the rms deviation obtained
with the standard choice $c/\sqrt{A}$ for the
average             
pairing-gap $\overline{\Delta}$ or effective-interaction
pairing gap  $\Delta_G$.
 
\item               
The Lipkin-Nogami pairing model yields
an rms deviation that is 14\% lower than the rms deviation in
the BCS approximation.
 
\item One cannot deduce the optimum constants for
a microscopic pairing model by simplified macroscopic
calculations.       
 
\item               
It is necessary to distinguish between several pairing-gap
concepts, notably the average pairing gap $\overline{\Delta}$, the
effective-interaction pairing gap $\Delta_G$
used as input in microscopic calculations,
and the microscopic 
pairing gap $\Delta$ obtained as a solution to the
BCS or Lipkin-Nogami pairing equations.
 
\item               
The effective-interaction pairing-gap $\Delta_G$ does not
depend explicitly on the relative neutron excess $I$.
\end{enumerate}

Our first four color figures represent results of theoretical
calculations of pairing quantities in the Lipkin-Nogami approximation.
In this model a pairing gap $\Delta$ and number-fluctuation constant
$\lambda_2$ are obtained as solutions of the pairing equations. It is
the sum $\Delta + \lambda_2$, which we denote by $\Delta_{\rm LN}$,
that should be compared to odd-even mass differences. This quantity is
shown in Fig.~\ref{lippdpl2} for protons.  The areas enclosed in jagged
black lines are regions where experimental pairing gaps can be
extracted from experimental masses by use of fourth-order
finite-difference expressions. Because magic numbers and the $N=Z$
Wigner cusp give non-smooth contributions to the mass surface and
corresponding contributions to the finite-difference expressions, it is
not possible to extract experimental pairing gaps from mass differences
in certain regions  
near magic numbers and near $N=Z$.$\,^{7,34})$\ess  The
proton pairing gap $\Delta_{\rm LN_{\rm p}}$ on the whole
decreases with increasing
$A$ and with increasing $Z$\@. There is a decrease
in the individual contribution $\Delta_{\rm p}$, which is not
shown or tabulated in this paper, at magic proton numbers. This
decrease is compensated for by a strong increase in $\lambda_{\rm 2p}$,
which results in a smooth appearance of $\Delta_{\rm LN_{\rm p}}$ at
magic proton numbers.
 
The  calculated neutron pairing gap $\Delta_{\rm LN_{\rm n}}$ is shown
in Fig.~\ref{lipndpl2}. The regions for neutrons where experimental pairing gaps can
be extracted from odd-even mass differences  are slightly
different from those for protons shown in Fig.~\ref{lippdpl2}.  The
individual contribution $\Delta_{\rm n}$ decreases considerably near
magic neutron numbers. This decrease is compensated for by a strong
increase in $\lambda_{\rm 2n}$, so that their sum behaves relatively
smoothly at magic neutron numbers, as seen in Fig.~\ref{lipndpl2}.  The
pairing gap $\Delta_{\rm LN_{\rm n}}$ on the whole decreases with increasing $A$ and
with increasing $N$.  There is no collapse at magic neutron numbers, in
contrast to results based on the BCS approximation.

As discussed in our previous pairing
study,$\,^{7})$\mss one may also determine pairing gaps directly
from odd-even mass differences based on {\it theoretical\/} masses.  This
type of theoretical pairing gap we denote by $\Delta_{\rm thmass}$.
There are several strong reasons to expect that it is more appropriate to
compare $\Delta_{\rm thmass}$ to $\Delta_{\rm exp}$ than to compare
$\Delta_{\rm LN}$ to $\Delta_{\rm exp}$.  The odd-even mass differences
pick up any non-smooth contributions to the nuclear-mass surface,
in addition to those represented by
the pairing gaps. Although we have excluded regions near magic
numbers from consideration, there are other gaps in the level spectra
that are expected to give non-smooth contributions in the odd-even mass
differences. These include the $N=56$ spherical subshell near
the proton gap at $Z=40$ and
the $N=152$  deformed gap near the proton
gap at  $Z=100$ in the actinide region.  Shape transitions
also give rise to non-smooth components to odd-even mass
differences. Such additional contributions to
$\Delta_{\rm exp}$ are expected near $N=88$ in the transition to the deformed
rare-earth region, for example.  Figures \ref{errpln} and \ref{errnln}
clearly exhibit discrepancies related to these gaps and shape transitions
that {\it cannot\/} automatically be interpreted as
an inherent deficiency of the microscopic pairing model.
 
These non-smooth contributions should be equally
present in both $\Delta_{\rm exp}$ and $\Delta_{\rm thmass}$,
provided that the theoretical calculation accurately describes the nuclear
mass surface. Although neither $\Delta_{\rm exp}$  nor $\Delta_{\rm
thmass}$ represent the true pairing gap because of these non-smooth
contributions, they should cancel
out in the discrepancy
$\Delta_{\rm exp} - \Delta_{\rm thmass}$.
In Table~\ref{tabpair1} we present a study of the two discrepancies
$\Delta_{\rm exp} - \Delta_{\rm LN}$ and
$\Delta_{\rm exp} - \Delta_{\rm thmass}$.
\begin{table}[t]    
\tabcolsep=0.077in  
\caption[tabpair1]{\baselineskip=12pt\small Errors of
\label{tabpair1} pairing-gap calculations. The experimental data base
contains 1654 masses.$\,^{2})$\ess}
\begin{center}      
\begin{tabular}{lrccccccccc}
\hline\\[-0.07in]   
& & \multicolumn{4}{c}{LN} &
&\multicolumn{4}{c}{Mass model} \\[0.08in]
\cline{3-6} \cline{8-11} \\[-0.07in]
Nucleons & \multicolumn{1}{c}{$N_{\rm nuc}$}
& $\mu_{\rm th}$& $\sigma_{\rm th}$
&  $\sigma_{\rm th;\mu=0}$  &   rms & &
$\mu_{\rm th}$ &    
$\sigma_{\rm th}$ &  $\sigma_{\rm th;\mu=0}$    & \multicolumn{1}{c}{rms} \\
  &       & (MeV) & (MeV) & (MeV)
 & (MeV) &  & (MeV) &(MeV) & (MeV) &\multicolumn{1}{c}{(MeV)} \\[0.08in]
\hline\\[-0.07in]   
Neutrons &  756 & 0.0220 & 0.1610 & 0.1627 &  0.1712 &  &  $-0.0427$
                                                & 0.1929 & 0.1972 & 0.2088\\
Protons &  648 & 0.0455 & 0.1529 & 0.1596 &  0.1670 & &  $-0.0096$
                                              & 0.1728 & 0.1730 & 0.1994 \\
Total   & 1404 & 0.0328 & 0.1577 & 0.1613 &  0.1692 & &  $-0.0274$
                                          & 0.1846 & 0.1865 & 0.2045\\[0.08in]
\hline              
\end{tabular}\\[3ex]
\end{center}        
\end{table}         
In contrast to our expectations, we find better agreement between
$\Delta_{\rm exp}$ and  $\Delta_{\rm LN}$  than between $\Delta_{\rm
exp}$ and  $\Delta_{\rm thmass}$.
The total second    
central moment is  16\% larger in the latter
comparison than in the former.  The reason
that we obtain such a large error for $\Delta_{\rm thmass}$ is that the
deviation of the calculated mass surface from the true mass surface is
sufficiently large to cancel the advantages of using $\Delta_{\rm
thmass}$ as a theoretical pairing gap instead of $\Delta_{\rm LN}$.
In the transition regions between spherical and
deformed nuclei near magic numbers, where we had expected such
advantages, the error in the calculated mass
surface is especially large, as can be seen in the comparison of
measured and calculated masses in Fig.~\ref{masd92a}.

\subsection{Total binding energy}
 
The total binding energy is equivalent to the atomic mass
excess, which has been discussed extensively in our mass paper.$\,^{1})$\ess
It is listed in the Table  to facilitate applications
that require  binding energy rather than mass excess.
 
\subsection{Neutron separation energies}
 
One- and two-neutron separation energies are shown versus $N$ and $Z$ in
Figs.~\ref{s1nod92a}--\ref{s2n92a}.  The $r$-process takes place
primarily in        
the region 2.0~MeV~$<S_{\rm 1n}<3.0$~MeV for odd $N$. Near $N=50$ and
$N=82$ the region of known nuclei extends into the $r$-process region.
 
The discrepancy between experimental one-neutron separation energies
obtained from mass differences and calculated one-neutron separation
energies is shown in Fig.~\ref{errs1n92a}. The biggest errors occur
near magic numbers. There is a general decrease of the error as $A$ increases.
 
\subsection{$\beta$-decay properties \label{betcal}}
 
A detailed knowledge of the low-energy part of the $\beta$-strength
function is essential for the calculation of many nuclear-structure
quantities of astrophysical interest, such as the probability of
$\beta$-delayed neutron and proton emission, the probability of
$\beta$-delayed fission, and half-lives with respect to $\beta$ decay.
Experimentally it has been known for some time that the low-energy part
of the $\beta$-strength function exhibits a pronounced
structure,$\,^{35,36})$\ess where the strength is collected
in a few well-localized peaks. For nuclei that are spherical in their
ground state there are usually only a few peaks within the $Q_{\beta}$
window; for deformed nuclei the strength is more spread out, but still
exhibits significant structure.
 
Theoretically, these main features of the $\beta$-strength function can be
understood on the basis of an extreme single-particle model.
The peaks in the strength functions correspond to transitions between
specific single-particle levels. In the spherical case the levels are
highly degenerate and spaced far apart, which gives rise to very
few but strong peaks in the experimental strength function. For
deformed nuclei the degeneracy is removed, allowing  for significantly
more transitions. Thus, compared to the spherical case, there are now
more peaks in the experimental strength function, but the strength of
each peak is lower. 
 
Although an extreme single-particle model explains the origin of the
structure in the $\beta$-strength function and the characteristic
difference between strength functions associated with deformed and
spherical nuclei, a more detailed description of the strength function
requires the consideration of the residual pairing and Gamow-Teller
interactions discussed above. The inclusion of these terms in the
potential reduces the calculated strength in the low-energy part of the
strength function to about 10\% of what is obtained in an extreme
single-particle model.$\,^{17})$\ess  Because pairing leads
to a diffuse Fermi surface and consequently to some occupation
probability above and to partially unfilled levels below the Fermi
surface, there are decay channels open in the more refined model that
are blocked in the extreme single-particle model.
For deformed nuclei one often finds considerable strength for
transitions between Nilsson levels whose asymptotic quantum numbers do
not allow for any transition probability according to the GT selection
rules.  This  occurs because the conventional, asymptotic
quantum-number label gives the main component of the wave function
corresponding to the level, but the transition strength is due to
small               
admixtures of wave functions with other asymptotic quantum numbers.
Since we perform a full diagonalization of the single-particle
Hamiltonian we account for these admixtures in our model.

The $\beta$-strength function is a sensitive measure of the
underlying single-particle structure.
Therefore,  when calculating
a $\beta$-strength function, two
conditions should ideally be fulfilled.  First, the ground-state shape of the system
of interest must be known. Second, the single-particle spectrum
calculated at this shape must agree reasonably well with the
experimental situation, especially for the levels closest to the
Fermi surface.  The calculation of a $\beta$-decay half-life also
requires the energy released in the decay, or
equivalently the ground-state masses of the mother and daughter
nuclei.  We now use the folded-Yukawa single-particle
potential, which has been applied to the calculation
of nuclear masses, shapes, and
other ground-state quantities for nuclei throughout the periodic
system,$\,^{1,33,37})$\mss
where these requirements are met globally.
We noted in Sec.~\ref{gspin}  that for
spherical nuclei there was excellent agreement between calculated and
measured ground-state spins and parities. For deformed regions
disagreements occurred somewhat more
often. However, when a disagreement occurs, a level with the correct spin and parity
is often calculated to be very near the last occupied orbital. If this
situation occurs in the daughter nucleus there is a rather small effect
on the calculated $\beta$-strength function. If, however,  in a decay
from an odd-even or odd-odd nucleus the unpaired proton or neutron is
in an incorrect level the effect can be more significant.
Therefore, it is reasonable to expect
characteristic differences in the discrepancies between calculated and
experimental half-lives in odd-even and even-even decay. Surprisingly,
as will be discussed below, we see no such differences.
 
To illustrate some major features associated with $\beta$-decay we show
in Figs.~\ref{bsrb95} and \ref{bsrb99} calculated $\beta$-strength
functions for the spherical nucleus $^{95}$Rb and the deformed nucleus
$^{99}$Rb.          
Although we obtain  
a fairly large  deformation for $^{95}$Rb
in our calculation of nuclear ground-state masses and deformations$\,^{1})$
we have performed the calculation shown in Fig.~\ref{bsrb95} for a spherical
shape, which is consistent with experimental information on this nucleus,
so that we can study $\beta$-strength-function
features associated with a sudden onset of deformation as the neutron number
increases from $N=58$ to $N=62$.
As discussed in more detail in Ref.$\,^{18})$, the single-particle
and pairing properties are evaluated for the appropriate vacuum nucleus.
In addition, the $Q_{\beta}$ value corresponds to the decay
indicated and the $S_{\nu {\rm n}}$ and  $P_{\nu {\rm n}}$ are evaluated for
the daughter of the decay.
A characteristic difference between $\beta$-strength
functions of spherical and deformed nuclei is that the former contain
only a few strong peaks in the low-energy region, whereas the latter
contain many more peaks of smaller size. Therefore, the likelihood that
one or several of them will occur close to zero energy is much larger
in the deformed case than in the spherical case.  Because the
transition rate is proportional to $S_{\beta}(E_x - Q_{\beta})^5$,
where $E_x$ is the excitation energy of the daughter state, there is
therefore usually a characteristic drop in the $\beta$-decay half-life
at the transition from a  spherical to a deformed system.

It is not our aim here to make a detailed analysis of each individual
nucleus, but instead to present an overview of the model performance in
a calculation of a large number of $\beta$-decay half-lives.  In
Figs.~\ref{betlifmn}--\ref{betlifpq} we compare measured and calculated
$\beta$-decay half-lives for nuclei throughout the periodic system.
The experimental half-lives are from a compilation by
Browne.$\,^{38})$\ess We present results with respect to
$\beta^-$ decay and also to $\beta^+$ decay and electron capture.  To
avoid lengthy constructions we will in our discussion in this section
usually not distinguish between $\beta^+$ decay and electron capture
and somewhat inexactly take $Q_{\beta}$ to mean the maximum energy
release in the decay.  Because the calculated pairing gap affects o-o,
o-e, and e-e decays differently we analyze these decays separately to
differentiate between effects due to pairing and due to other causes.
 
We have limited the comparison  to nuclei whose experimental half-lives
are shorter than 1000~s. Because the relative error in the calculated
half-lives is more sensitive to small shifts in the positions of the
calculated single-particle levels for decays with small energy releases,
where long half-lives are expected, one can anticipate that
half-life calculations are more reliable far from stability than close
to $\beta$-stable nuclei. To address the reliability in various regions
of nuclei and versus distance from stability, we present in
Figs.~\ref{betlifmn}--\ref{betlifpq} the ratio $T_{\beta,{\rm
calc}}/T_{\beta,{\rm exp}}$ versus the {\it three\/} different
quantities $N$, $T_{\beta,{\rm exp}}$, and $Q_{\beta}$.
 
The few cases that lie outside the scale of the figures have  been
taken into account in the error analysis presented below.  These cases
are associated with long experimental half-lives; there is only one
case outside the scale of the figure
with an experimental half-life less than 10~s. In such cases
first-forbidden transitions, which we do not take into account at this
stage, can be expected to make a considerable contribution to the decay
rate.  Therefore, one should not conclude from these relatively few
very large deviations that the calculated {\it Gamow-Teller\/}
$\beta$-strength function is in serious disagreement with experimental
data.               
 
Before we make a quantitative analysis of the agreement between
calculated and experimental half-lives we briefly discuss what
conclusions can be drawn from a simple visual
inspection of Figs.~\ref{betlifmn}--\ref{betlifpq}.
In Figs.~\ref{betlifmn} and \ref{betlifpn}
the quantities $T_{\beta,{\rm calc}}/T_{\beta,{\rm exp}}$
are plotted as functions of neutron number $N$. There are no
systematic trends with $N$. For $\beta^-$ decay of
even-even nuclei the calculated
half-lives are somewhat too long on the average.
In Figs.~\ref{betlifmt} and \ref{betlifpt} the quantities
$T_{\beta,{\rm calc}}/T_{\beta,{\rm exp}}$ are plotted as functions of
the experimental half-life $T_{\beta,{\rm exp}}$. As a function of this
quantity, one would expect the average error to increase as
$T_{\beta,{\rm exp}}$ increases.  This is indeed the case for $\beta^-$
decay, but not for $\beta^+$ decay, except for odd-odd decays---an
unexpected result.  
In Figs.~\ref{betlifmq} and \ref{betlifpq} the quantities
$T_{\beta,{\rm calc}}/T_{\beta,{\rm exp}}$ are plotted as functions of
$Q_{\beta}$ with the aim of showing how the average error increases as
$Q_{\beta}$ decreases. It is obvious that errors in the location of the
peaks in the calculated strength function have a larger effect on the
calculated half-lives for small values of $Q_{\beta}$  than for larger
ones.  Indeed, we see a fairly clear increase in the scatter of the
points in Fig.~\ref{betlifmq} as $Q_{\beta}$ decreases. However, for
$\beta^+$ decay shown in Fig.~\ref{betlifpq} the only correlation
occurs for o-o decay.

In a visual inspection of Figs.~\ref{betlifmn}--\ref{betlifpq} one is
left with the impression that the errors in our calculation are fairly
large. However, this is partly a fallacy, since for small errors there
are many more points than for large errors. This is not clearly seen in
the figures, since for small errors many points are superimposed on
one another. To obtain a more exact understanding of the error in the
calculation we therefore perform a more detailed analysis.

One often analyzes the error in a calculation by studying a
root-mean-square deviation, which in this case would be
\beq                
{\sigma_{\rm rms}}^2 = \frac{1}{n}\sum_{i=1}^{n} (T_{\beta,{\rm exp}} -
T_{\beta,{\rm calc}})^2
\eeq{sigrms}        
However, such an error analysis is
unsuitable here, for two reasons.  First, the quantities studied vary
by many orders of magnitude. In our case the variation is more than ten
orders of magnitude, from the millisecond range to years and beyond.
Second, the calculated  and measured quantities may {\it differ} by orders of
magnitude. We therefore study the quantity $\log
(T_{\beta,{\rm calc}}/T_{\beta,{\rm exp}})$, which is plotted  in
Figs.~\ref{betlifmn}--\ref{betlifpq},
instead of $(T_{\beta,{\rm exp}} - T_{\beta,{\rm calc}})^2$.
 
To facilitate the interpretation of the error plots we consider two
hypothetical cases.  As the first example, suppose that all the points
were grouped on the line $T_{\beta,{\rm calc}}/T_{\beta,{\rm
exp}}=10$.  It is immediately clear that an error of this type could be
entirely removed by introducing a  renormalization factor, which is a
common practice in the calculation of $\beta$-decay half-lives.  We
shall see below that in our model the half-lives corresponding to our
calculated  strength functions have about zero average deviation from
the calculated half-lives, so no renormalization factor is necessary.
 
In another extreme, suppose half the
points were located on the line
$T_{\beta,{\rm calc}}/T_{\beta,{\rm exp}}=10$
and the other half on the line
$T_{\beta,{\rm calc}}/T_{\beta,{\rm exp}}=0.1$.
In this case the average of
$\log(T_{\beta,{\rm calc}}/T_{\beta,{\rm exp}})$
would be zero.  We are
therefore led to the conclusion that there are two types of errors that
are of interest to study, namely the average position of the points in
Figs.~\ref{betlifmn}--\ref{betlifpq},
which is just the average of the quantity
$\log(T_{\beta,{\rm calc}}/T_{\beta,{\rm exp}})$,
and the spread of the points around this
average.  To analyze the error along these ideas, we
introduce the quantities
\beqar              
                 r &=&T_{\beta,{\rm calc}}/T_{\beta,{\rm exp}} \nonumber \\[1ex]
                 r_{\rm l} &=& \log_{10}(r) \nonumber \\[1ex]
                 M_{r_{\rm l}} &=&
                    \frac{1}{n}\sum_{i=1}^n r_{\rm l}^i \nonumber \\[1ex]
                 M_{r_{\rm l}}^{10} &=& 10^{M_{r_{\rm l}}}   \nonumber \\[1ex]
\sigma_{r_{\rm l}} &=& {\left[
\frac{1}{n} \sum_{i=1}^n {\left( r_{\rm l}^i
- M_{r_{\rm l}} \right)}^2 \right] }^{1/2} \nonumber \\[1ex]
\sigma_{r_{\rm l}}^{10} &=& 10^{\sigma_{r_{\rm l}}}
\eeqar{statdef}     
where $M_{r_{\rm l}}$ is the average position of the points
and $\sigma_{r_{\rm l}}$ is the spread around this average.
The spread $\sigma_{r_{\rm l}}$ can be expected to be related
to uncertainties in the positions of the levels in the
underlying single-particle model.
The use of a logarithm in the definition of
$r_{\rm l}$ implies that these two
quantities correspond directly to distances as seen by the eye in
Figs.~\ref{betlifmn}--\ref{betlifpq},
in units where one order of magnitude is 1. After the error
analysis has been carried out we want to discuss its result
 in terms like ``on the average the calculated half-lives
are `a factor of two' too long.'' To be able to do this we must convert
back from the logarithmic scale. Thus, we realize
that the quantities $M_{r_{\rm l}}^{10}$ and $\sigma_{r_{\rm l}}^{10}$
are conversions back to ``factor of'' units of
the quantities $M_{r_{\rm l}}$ and $\sigma_{r_{\rm l}}$,
which are expressed in distance
or logarithmic units.
 
In Table \ref{tabbetm}
\begin{table}[b]    
\caption[betlifmt]  
{Analysis of the discrepancy between calculated and \label{tabbetm}
measured $\beta^-$-decay half-lives shown in Fig.~\ref{betlifmt}.}
\begin{center}      
\begin{tabular}{rrrrrrr}
\hline\\[-0.07in]   
& $ n $ & $M_{r_{\rm l}}$ & $ M_{r_{\rm l}}^{10} $ & $  \sigma_{r_{\rm l}}$
 & $\sigma_{r_{\rm l}}^{10}$ & $T_{\beta,{\rm exp}}^{\rm max}$\\
&  & & &  & & (s)\\[0.08in]
\hline\\[-0.07in]   
o-o &   29&  $-0.23$&   0.59&   0.46&   2.91&     1\\
o-e &   35&  $-0.23$&   0.59&   0.42&   2.64&     1\\
e-e &   10&     0.58&   3.84&   0.49&   3.08&     1\\[1ex]
o-o &   59&  $-0.12$&   0.76&   0.95&   8.83&    10\\
o-e &   85&  $-0.11$&   0.78&   0.68&   4.81&    10\\
e-e &   34&     0.40&   2.50&   0.62&   4.13&    10\\[1ex]
o-o &   88&     0.37&   2.33&   1.69&  49.19&   100\\
o-e &  133&     0.05&   1.11&   0.98&   9.45&   100\\
e-e &   54&     0.42&   2.61&   0.68&   4.75&   100\\[1ex]
o-o &  115&     0.54&   3.50&   1.86&  72.02&  1000\\
o-e &  194&     0.44&   2.77&   1.85&  71.50&  1000\\
e-e &   71&     0.84&   6.86&   1.77&  58.48&  1000\\[1ex]
\hline              
\end{tabular}       
\end{center}        
\end{table}         
we show the results of an evaluation of the quantities in
Eq.~(\ref{statdef}) for $\beta^-$ decay of nuclei
with $T_{\beta,{\rm exp}} \leq  1000$~s.
For long            
half-lives one  expects forbidden decay to dominate.
We find that        
in the o-o and o-e  cases
the value of $M_{r_{\rm l}}^{10}$
increases from values somewhat below  1
to values somewhat above 1 as the maximum
half-life, $T_{\beta,{\rm exp}}^{\rm max}$,
considered increases from 1~s to 1000~s.
When cases corresponding to
long half-lives contribute to the error estimates,
first-forbidden decays would
be expected to increase the $\beta$-decay  rates. In our
model we neglect, at this stage, first-forbidden
decay. Therefore, one can expect that
as $T_{\beta,{\rm exp}}^{\rm max}$ increases,
our calculated half-lives
will become increasingly longer than experiment. This is
indeed the trend in Table~\ref{tabbetm}. Therefore,
it is preferable to use only nuclei with short half-lives to
determine if any renormalization of the $\beta$-strength function
is needed to reproduce on the average the experimental half-lives.
However, for a short half-life cutoff, for example
1~s,  there are not enough data
points for a reliable determination. Slightly higher half-life cutoffs
indicate $ M_{r_{\rm l}}^{10} \approx 1, $ except for the e-e case. However,
there are only a few data points for the e-e case, so the higher
 $M_{r_{\rm l}}^{10} $ obtained in this case may be unreliable,
especially in view of the results obtained below for $\beta^+$ decay.

In Table \ref{tabbetp}
we show the results of an evaluation of the quantities in
Eq.~(\ref{statdef}) for $\beta^+$ decay and electron capture  of nuclei
with $T_{\beta,{\rm exp}} \leq  1000$~s.
\begin{table}[b]    
\caption[betlifpt]  
{Analysis of the discrepancy between calculated and \label{tabbetp}
measured $\beta^+$-decay
and electron-capture half-lives shown in Fig.~\ref{betlifpt}.}
\begin{center}      
\begin{tabular}{rrrrrrr}
\hline\\[-0.07in]   
& $ n $ & $M_{r_{\rm l}}$ & $ M_{r_{\rm l}}^{10} $ & $  \sigma_{r_{\rm l}}$
 & $\sigma_{r_{\rm l}}^{10}$ & $T_{\beta,{\rm exp}}^{\rm max}$\\
&  & & &  & & (s)\\[0.08in]
\hline\\[-0.07in]   
o-o &    21&     0.17&   1.49&   0.60&   3.99&     1\\
o-e &    30&     0.25&   1.79&   0.60&   3.97&     1\\
e-e &     9&     0.55&   3.52&   0.31&   2.03&     1\\[1ex]
o-o &    43&     0.09&   1.22&   0.76&   5.77&    10\\
o-e &    77&     0.03&   1.07&   0.53&   3.38&    10\\
e-e &    33&     0.21&   1.62&   0.65&   4.46&    10\\[1ex]
o-o &    85&     0.11&   1.30&   1.06&  11.37&   100\\
o-e &   149&  $-0.14$&   0.73&   0.52&   3.33&   100\\
e-e &    63&  $-0.01$&   0.98&   0.55&   3.52&   100\\[1ex]
o-o &   146&     0.14&   1.37&   1.24&  17.49&  1000\\
o-e &   238&  $-0.20$&   0.63&   0.65&   4.47&  1000\\
e-e &   101&  $-0.08$&   0.83&   0.50&   3.16&  1000\\[1ex]
\hline              
\end{tabular}       
\end{center}        
\end{table}         
Above we argued in the case of $\beta^-$ decay that as nuclei with
longer half-lives are included, one can expect the calculated
half-lives to be too long because the rates due to first-forbidden
decay are not included.  However, we see  instead
a slightly opposite trend here. But from arguments similar to those used
in the discussion of $\beta^-$ decay we conclude that in $\beta^+$
decay  our results are consistent with zero average deviation between
calculated and experimental half-lives. Thus, the error in the
calculation is given by $\sigma_{r_{\rm l}}^{10}$ only. The behavior of
$\sigma_{r_{\rm l}}^{10}$  as $T_{\beta,{\rm exp}}^{\rm max}$ increases
is similar for both $\beta^-$ and $\beta^+$ decay. Just as when the
average deviation is determined, one should not consider decays with
too-long half-lives when determining $\sigma_{r_{\rm l}}^{10}$.
On the other hand, it would be desirable to base the determination
of $\sigma_{r_{\rm l}}^{10}$  on a large set of data. Both
of these conditions cannot   be well-fulfilled simultaneously.
However, if we      
base our estimate of the model error $\sigma_{r_{\rm l}}^{10}$
by weighting most heavily  half-lives in the shorter range, we
find that the model error is about a factor of four.
 
Certain additional assumptions and approximations$\,^{18})$ have
been made to extend the \mbox{odd-$A$} QRPA formalism to
the odd-odd case. If these simplifications
were inadequate, one would expect to see a larger $\sigma_{r_{\rm l}}$ in the
o-o case than in the o-e case. No clear effect of this type is seen in
Table~\ref{tabbetp} for the shorter half-lives but
does develop for very long experimental half-lives.
 
We have already mentioned that particle-particle
correlations are not taken into account in our QRPA treatment.  These
correlations are expected to strongly suppress $\beta^+$
transitions,$\,^{21,39})$\mss and as a consequence lead to longer
half-lives than those obtained in a model that does not consider
these correlations. However, our results in
Figs.~\ref{betlifmn}--\ref{betlifpq} and Tables \ref{tabbetm} and
\ref{tabbetp} show that  in a standard treatment with
$\chi_{\rm GT}= 23$ MeV$/A$ and no renormalization coefficient,
we obtain $\beta^+$-decay rates that actually agree better with experiment
than do our calculated $\beta^-$-decay rates.
 
\subsection{Proton separation energies}
 
One- and two-proton separation energies are
shown versus $N$ and $Z$ in Figs.~\ref{s1pod92a}--\ref{s2p92a}.
For odd $Z$ the region of known nuclei extends into
the region where proton emission is energetically allowed
in several places, but especially just above $N=82$.
However, the Coulomb barrier
severely inhibits   
proton emission from the ground state     
when the energy released is small.
Proton emission is therefore
more readily observed as delayed emission following
$\beta$ decay.$\,^{40,41})$\ess

The discrepancy between experimental one-proton separation
energies obtained from mass differences and calculated one-proton
separation energies is shown in Fig.~\ref{errs1p92a}. The
biggest errors occur near the magic numbers. There is a general
decrease of the error as $A$ increases.
 
\subsection{$\alpha$-decay properties}
 
The calculated energy release and associated
half-life  with respect to $\alpha$ decay are
plotted in Figs.~\ref{qal92a} and
\ref{talp92a}, respectively.
 
In 1989 M\"{u}nzenberg {\it et al\/}.$\,^{42})$
compared the $\alpha$-decay energy release
along the $N=154$ and 155 isotonic
lines to predictions of the  FRLDM (1988).$\,^{43})$\ess
In                  
Fig.~\ref{qcomp1} we make a similar comparison of measured data to
predictions of the current FRDM (1992).$\,^{1})$\ess These
results based on the current FRDM (1992) agree much better with the measured
values than do the corresponding results calculated from our older mass
model in 1989 by M\"{u}nzenberg {\it et al\/}.,$\,^{42})$\ess and
also much better than the results obtained with the 1991 version
of the FRDM.$\,^{44})$\ess  The improvement is due  partly to the
inclusion of the $\epsilon_6$ shape degree of freedom.

From the Table we find that the nucleus $^{272}$110 has a calculated
$\alpha$-decay half-life of  71~ms.  The nuclei $^{288}$110 and
$^{290}$110 in the center of the superheavy island have calculated
$\alpha$-decay half-lives of 4 y and 1565 y, respectively. With the
1991 version of the FRDM we obtained calculated $\alpha$-decay half-lives
of 16 ms, 161 y, and 438 y for $^{272}$110, $^{288}$110, and $^{290}$110,
respectively. Our current results are different from these older
ones for two reasons.
First,              
we now use the new mass model FRDM (1992).
Second,             
we now use a new set of constants in
Eq.~(\ref{violas}). 
For  $^{288}$110 the value of $Q_{\alpha}$ is 7.36 MeV
in our current model, but only 6.95 MeV in the FRDM (1991).  For
$^{272}$110 the difference in $Q_{\alpha}$ values is only 0.04
MeV\@,  and the difference in the calculated half-lives is in this case
due almost entirely to the new set of constants used in
Eq.~(\ref{violas}). 
 
In the introduction to Sec.~\ref{sec3} we pointed out that because the calculated
potential-energy surfaces for some very proton-rich nuclei are very flat
and contain multiple minima, the ground-state deformation may exhibit large
fluctuations between neighboring nuclei. For example, we see in
Ref.$\,^{1})$ that  the calculated quadrupole deformation of
$^{318}$128 is $\epsilon_2=0.0$ whereas  $\epsilon_2=0.392$
is obtained for $^{319}$128. Such deformation changes are
accompanied by non-smooth changes in calculated energy releases
and associated half-lives.
An $\alpha$-decay half-life of $10^{-8.15}$~s is obtained for
$^{318}$128, whereas the calculated $\alpha$-decay half-life for
$^{319}$128 is greater than $10^{20}$~s. Because of very short
spontaneous-fission half-lives for such proton-rich nonspherical nuclei,
we do not           
expect these nuclei to be observable despite their calculated
long $\alpha$-decay half-lives.
 
\section{EXTRAPOLATEABILITY \label{reli}}
 
Theories are often characterized by their rms error with respect to
experimental        
data points.        
However, this is often unsuitable because the rms error contains
contributions from errors in the experimental data points and is
therefore always larger than the intrinsic error of the
theory. On the other hand,
it is possible to determine the
intrinsic error of a theory in a way that contains no contributions
from experimental errors.$\,^{43})$\ess  It is most natural to
characterize the deviation between measured data points and theoretical
calculations  by two quantities: (1)
the mean deviation $\mu_{\rm th}$ of the theory from
the  data points and (2) the  standard deviation
$\sigma_{\rm th}$ of the theoretical results
about this mean.$\,^{1,45,46})$\ess
Alternatively, one may characterize
the error of a theory by a single number, the second central moment
of the error term of the theory,
which we  denote by $\sigma_{\rm th;\mu=0}$.
This measure is similar to the rms
deviation, but without the deficiency of contributions from
experimental errors.
We use  these four error measures in
our discussions  of model accuracies below.
 
After we completed our mass model,$\,^{1})$\mss a new compilation
of experimental ground-state masses$\,^{3,4})$ has become
available. We are therefore now able to compare our FRDM (1992) predictions
to data that were not taken into account when the model constants were
determined, instead of studying the model reliability through simulation.
In Table~\ref{tabfrdm} we give errors for the FRDM(1992) for
ground-state masses and six separation energies and energy releases
\begin{table}[b]    
\begin{small}       
\begin{center}      
\caption[tabmas]{Theoretical errors \label{tabfrdm} and
extrapolateability of the FRDM (1992).\\[2ex]}
\vspace{10.862pt}   
\begin{tabular}{lrccccrccc}
\hline\\[-0.07in]   
& \multicolumn{3}{c}{Original nuclei (1989)} & &
 \multicolumn{5}{c}{New nuclei (1989--1993)} \\[0.08in]
 \cline{2-4} \cline{6-10} \\[-0.07in]
 & $N_{\rm nuc}$    
 & $\sigma_{\rm th;\mu=0}$
 & rms              
 &                  
 & $N_{\rm nuc}$    
 & \multicolumn{1}{c}{$\mu_{\rm th}$}
 & $\sigma_{\rm th}$
 & $\sigma_{\rm th;\mu=0}$
 & rms              
\\                  
 & & (MeV) & (MeV) & & & (MeV) &
 \multicolumn{1}{c}{(MeV)} & (MeV) & (MeV) \\[0.08in]
\hline \\[-0.07in]  
$M_{\rm th}$        
               & 1654  & 0.669     & 0.681   &
               & 217  & $0.074$   & 0.638   & 0.642   &  0.730
\\[3pt]             
$S_{\rm 1n}$        
               & 1464  &  0.393   & 0.411   &
               &  195  &  0.033   & 0.525   & 0.526  & 0.624
\\[3pt]             
$S_{\rm 2n}$        
               & 1403  &  0.539  & 0.555   &
               & 198   & 0.026   & 0.575   & 0.576  &  0.684
\\[3pt]             
$Q_{{\beta^-}}$     
               & 1353  & 0.488   & 0.507   &
               &  216  &  0.006  & 0.559   & 0.559 &  0.647
\\[3pt]             
$S_{\rm 1p}$        
               & 1400  & 0.381   & 0.397   &
               &  210  & 0.068   & 0.576   & 0.581  &  0.672
\\[3pt]             
$S_{\rm 2p}$        
               & 1314  & 0.493   & 0.509   &
               &  193  & 0.026   & 0.585   & 0.586  & 0.663
\\[3pt]             
$Q_{\alpha}$        
 
               & 1450  &  0.625  & 0.641   &
               &  224  &  0.021  & 0.675   & 0.675  &  0.772
\\[0.08in]          
\hline              
\end{tabular}\\[3ex]
\end{center}        
\end{small}         
\end{table}         
for both the region in which the model constants were determined and
\begin{table}[tb]   
\tabcolsep = 0.036in
\begin{small}       
\begin{center}      
\caption[tabmas]{Comparison of theoretical errors and
extrapolateability of the FRDM (1992) with \label{tabetfsi} those of the
ETFSI-1 (1992) model.\\}
\begin{tabular}{lrcccccccrcccrcc}
\hline\\[-0.07in]   
& \multicolumn{6}{c}{Original nuclei (1989)} & &
 \multicolumn{8}{c}{New nuclei (1989--1993)} \\[0.08in]
 \cline{2-7} \cline{9-16} \\[-0.07in]
& & \multicolumn{2}{c}{FRDM} & &
\multicolumn{2}{c}{ETFSI-1} & & &
\multicolumn{3}{c}{FRDM} & & \multicolumn{3}{c}{ETFSI-1} \\[0.08in]
 \cline{3-4} \cline{6-7} \cline{10-12} \cline{14-16} \\[-0.07in]
 & \multicolumn{1}{c}{$N_{\rm nuc}$}
 & $\sigma_{\rm th;\mu=0}$
 & rms              
 &                  
 & $\sigma_{\rm th;\mu=0}$
 & rms              
 &                  
 & $N_{\rm nuc}$    
 & \multicolumn{1}{c}{$\mu_{\rm th}$}
 & $\sigma_{\rm th;\mu=0}$
 & rms              
 &                  
 & \multicolumn{1}{c}{$\mu_{\rm th}$}
 & $\sigma_{\rm th;\mu=0}$
 & rms              
\\                  
 & & (MeV) & (MeV) & & (MeV) & (MeV) &
 & & (MeV) & \multicolumn{1}{c}{(MeV)} & (MeV) & &
 (MeV) & \multicolumn{1}{c}{(MeV)} & (MeV) \\[0.08in]
\hline \\[-0.07in]  
$M_{\rm th}$        
& 1540              
& 0.607             
& 0.615             
&                   
& 0.733             
& 0.742             
&                   
& 213               
& $0.070$           
& 0.645             
& 0.716             
&                   
& $0.015$           
& 0.809             
& 0.866             
\\[3pt]             
$S_{\rm 1n}$        
& 1356              
& 0.335             
& 0.346             
&                   
& 0.514             
& 0.521             
&                   
& 190               
& 0.033             
& 0.525             
& 0.609             
&                   
& 0.044             
& 0.612             
& 0.681             
\\[3pt]             
$S_{\rm 2n}$        
& 1301              
& 0.477             
& 0.482             
&                   
& 0.501             
& 0.508             
&                   
& 192               
& 0.028             
& 0.580             
& 0.659             
&                   
& $-0.052$          
& 0.576             
& 0.659             
\\[3pt]             
$Q_{{\beta^-}}$     
& 1259              
& 0.425             
& 0.436             
&                   
& 0.679             
& 0.684             
&                   
& 212               
& 0.001             
& 0.560             
& 0.646             
&                   
& $-0.003$          
& 0.741             
& 0.791             
\\[3pt]             
$S_{\rm 1p}$        
& 1296              
& 0.324             
& 0.339             
&                   
& 0.466             
& 0.472             
&                   
& 205               
& 0.073             
& 0.581             
& 0.660             
&                   
& 0.113             
& 0.608             
& 0.709             
\\[3pt]             
$S_{\rm 2p}$        
& 1217              
& 0.434             
& 0.444             
&                   
& 0.520             
& 0.531             
&                   
& 189               
& 0.030             
& 0.589             
& 0.662             
&                   
& 0.082             
& 0.586             
& 0.674             
\\[3pt]             
$Q_{\alpha}$        
& 1342              
& 0.541             
& 0.546             
&                   
& 0.520             
& 0.529             
&                   
& 217               
& 0.018             
& 0.670             
& 0.762             
&                   
& 0.004             
& 0.618             
& 0.699             
\\[0.08in]          
\hline              
\end{tabular}\\[3ex]
\end{center}        
\end{small}         
\end{table}         
in the new region of nuclei.
 
We see in Table~\ref{tabfrdm} that the error $\sigma_{\rm th;\mu=0}$
for new masses, 0.642 MeV\@, is 4\%\ {\it smaller\/} than the error 0.669 MeV
in the region where the model constants were determined. Thus, we
conclude, as was done in our mass paper$\,^{1})$ through
simulations, that the model is very well-behaved as one moves away from
stability to new regions of nuclei. Indeed, at this stage there is no
sign that the model diverges.  {\it It is also very clear that it is not
possible to use the rms deviation as a proper measure of model error.}
It is 7\% larger in the new region of nuclei, although we have just
shown that the true model error is actually 4\%\ smaller in the new region
of nuclei than in the region where the model constants were determined.
This is because the experimental errors in the new region of
nuclei are quite large and contribute significantly to the rms error,
whereas they are normally quite small in the region of previously known
nuclei.             
 
In the absence of correlations in the model error between neighboring
masses, the model error for the separation energies and energy releases
would be  $\sqrt{2}$ times the mass error. However, Table~\ref{tabfrdm}
shows that the separation-energy and energy-release errors are instead
somewhat {\it smaller} than the mass-model error. This is
due to the correlation in the mass-model error for neighboring nuclei.
The separation-energy and energy-release errors increase in the new
region of nuclei, whereas the mass-model error does not.  This
indicates that the correlation between mass-model errors for
neighboring nuclei decreases in the new region. Since a mass model
should not be judged on the amount of error correlations between nearby
nuclei, one cannot gain significant insight about a model by comparing
the error of the separation energies and energy releases in the region
where the model constants were determined to the errors in a new region
of nuclei. However, because many of these quantities are used in
astrophysical calculations, there is a practical need to know the
errors of these quantities, and it is for this purpose that we provide
them.

In Table~\ref{tabetfsi}
we compare the errors of the FRDM (1992) and the
1992 extended Thomas-Fermi Strutinsky-integral model (version 1)
of Aboussir, Pearson, Dutta, and Tondeur$\,^{47,48})$
in the region of nuclei
where the ETFSI-1 (1992) model constants were
determined and in the region of new nuclei
that are included in the ETFSI-1 (1992) calculation.
The region of  nuclei considered in the ETFSI-1 (1992) model is
slightly different from the region considered in the FRDM (1992).
Its constants were also determined from
an adjustment to a slightly earlier
mass evaluation.$\,^{49})$\ess In Table~\ref{tabetfsi} the
quantities pertaining to the FRDM (1992) are also evaluated for this
more limited region. Therefore, the results for the FRDM (1992) in
Table~\ref{tabetfsi} are slightly different from those in
Table~\ref{tabfrdm}. The finalized ETFSI-1 (1992) mass table, as it
appears in Ref.$\,^{48})$, differs slightly from an earlier
table circulated privately before publication. We use here the
finalized, published version.
 
The comparison in Table~\ref{tabetfsi} between the extrapolateability
of the FRDM (1992) and the ETFSI-1 (1992) model  shows that for
these particular regions the FRDM (1992) error increases
by 6\% and the ETFSI-1  (1992) model error increases
by 10\% in the new region relative
to the region where the ETFSI-1 (1992) model constants were
determined.  Thus, we conclude from Tables \ref{tabfrdm} and \ref{tabetfsi}
that the FRDM (1992) is somewhat more stable
as one moves away from stability than is the ETFSI-1 (1992) model.

In Table~\ref{tabgroote} we compare the extrapolateability
\begin{table}[tb]   
\tabcolsep = 0.0185in
\begin{small}       
\begin{center}      
\vspace{-12pt}      
\caption[tabmas]{Comparison of theoretical errors and
\label{tabgroote} extrapolateability of the FRDM (1992/1977) with those of the
mass formula of von Groote {\it et al.\/} (1976).\\}
\begin{tabular}{lrcccccrccrcccrccrc}
\hline\\[-0.07in]   
& \multicolumn{4}{c}{Original nuclei (1977)}   & &
  \multicolumn{6}{c}{New nuclei (1977--1989)} & &
  \multicolumn{6}{c}{New nuclei (1989--1993)} \\[0.08in]
 \cline{2-5} \cline{7-12} \cline{14-19} \\[-0.07in]
& &   \multicolumn{1}{c}{FRDM} & &
\multicolumn{1}{c}{v.\ Groote} & & &
\multicolumn{2}{c}{FRDM} & & \multicolumn{2}{c}{v.\ Groote} & & &
\multicolumn{2}{c}{FRDM} & & \multicolumn{2}{c}{v.\ Groote} \\[0.08in]
 \cline{3-3} \cline{5-5} \cline{8-9} \cline{11-12}
                           \cline{15-16} \cline{18-19} \\[-0.07in]
 & \multicolumn{1}{c}{$N_{\rm nuc}$}
 & $\sigma_{\rm th;\mu=0}$
 &                  
 & $\sigma_{\rm th;\mu=0}$
 &                  
 & $N_{\rm nuc}$    
 & \multicolumn{1}{c}{$\mu_{\rm th}$}
 & $\sigma_{\rm th;\mu=0}$
 &                  
 & \multicolumn{1}{c}{$\mu_{\rm th}$}
 & $\sigma_{\rm th;\mu=0}$
 &                  
 & $N_{\rm nuc}$    
 & \multicolumn{1}{c}{$\mu_{\rm th}$}
 & $\sigma_{\rm th;\mu=0}$
 &                  
 & \multicolumn{1}{c}{$\mu_{\rm th}$}
 & $\sigma_{\rm th;\mu=0}$
\\                  
 &  &  (MeV)         & & (MeV) &
 &  &  \multicolumn{1}{c}{(MeV)} & (MeV) & &
                          \multicolumn{1}{c}{(MeV)} & (MeV) &
 &  &  \multicolumn{1}{c}{(MeV)} & (MeV) & &
                          \multicolumn{1}{c}{(MeV)} & (MeV)  \\[0.08in]
\hline \\[-0.07in]  
$M_{\rm th}$  & 1323 & 0.671 &
                       & 0.629 &
                &  351 & $0.004$ & 0.686 &
                       & $0.612$ & 1.154 &
                &  217 & $0.030$ & 0.668 &
                       & $0.808$ & 1.284 \\[3pt]
$S_{\rm 1n}$    &  1139 & 0.391 &
                        & 0.383 &
                &  335  & $-0.032$ & 0.406 &
                        & $-0.107$ & 0.431 &
                &  195  & 0.024    & 0.525 &
                        & $-0.095$ & 0.541    \\[3pt]
$S_{\rm 2n}$    &  1092 & 0.508 &
                        & 0.480 &
                &  320 & $-0.053$ & 0.632 &
                       & $-0.171$ & 0.650 &
                &  198 & 0.007    & 0.576 &
                       & $-0.213$ & 0.609    \\[3pt]
$Q_{{\beta^-}}$   & 1052  & 0.467 &
                          & 0.516 &
                &  314 & 0.077 & 0.552 &
                       & 0.193 & 0.737 &
                &  216 & 0.031 & 0.562 &
                       & 0.178 & 0.633    \\[3pt]
$S_{\rm 1p}$    & 1105  & 0.361 &
                        & 0.412 &
                &  305 & 0.039 & 0.463 &
                       & 0.102 & 0.548 &
                &  210 & 0.083 & 0.585 &
                       & 0.157 & 0.619    \\[3pt]
$S_{\rm 2p}$    & 994  & 0.478 &
                       & 0.501 &
                & 331  & 0.049 & 0.546 &
                       & 0.160 & 0.691 &
                & 193  & 0.056 & 0.591 &
                       & 0.217 & 0.738    \\[3pt]
$Q_{\alpha}$    &  1144 & 0.627  &
                        & 0.525  &
                & 321  & $-0.005$ & 0.630 &
                       & $-0.001$ & 0.669 &
                & 224  & 0.011    & 0.675 &
                       & 0.024    & 0.674    \\[0.08in]
\hline              
\end{tabular}\\[3ex]
\end{center}        
\end{small}         
\end{table}         
between the FRDM (1992/1977) and the 1976 mass formula of
von Groote, Hilf, and Takahashi.$\,^{50})$\ess
In this comparison we have adjusted the constants of the FRDM  to
similar masses$\,^{51})$ as those
considered by von Groote {\it et al.\/}\
in the determination of their constants. Clearly,
the mass formula of von Groote {\it et al.\/}\  diverges severely in
new regions of nuclei.
 
In Figs.~\ref{devmfrdm}--\ref{devqbmgro}
we plot the deviations between calculated and experimental
masses, one-neutron separation energies, and energy releases for
$\beta^-$ decay for these three models. Tables~\ref{tabfrdm}--\ref{tabgroote}
and Figs.~\ref{devqbmfrdm}--\ref{devqbmgro} are based on
{\it all\/} $Q_{\beta^{-}}$ values, both positive and negative.
Neither the FRDM (1992) nor the ETFSI-1 (1992) model shows
large divergences in the new region, but strong
systematic deviations occur in the model of
 von Groote {\it et al.\/}
 
We have also made a limited study of the extrapolateability of
the recent 1994 Thomas-Fermi model of Myers and
Swiatecki.$\,^{52,53})$\ess
In this model, the macroscopic energy is
calculated for a generalized Seyler-Blanchard nucleon-nucleon
interaction by use of the original Thomas-Fermi approximation.
For $N,Z\geq 30$ the shell and pairing corrections were taken from the 1992
finite-range droplet model, and for
$N,Z \leq 29$ a semi-empirical expression was used. The constants of the
model were determined by an adjustment to the ground-state masses of the same
1654 nuclei with $N,Z \geq 8$ ranging from $^{16}$O to $^{263}$106
whose masses were known experimentally in 1989 that were used in the
1992 finite-range droplet model. The theoretical error
corresponding to these 1654 nuclei is 0.640 MeV\@.
The reduced theoretical error relative to that in the 1992 finite-range
droplet model arises primarily from the use of semi-empirical
microscopic corrections in the region $N,Z \leq 29$ rather than microscopic
corrections calculated more fundamentally.
The theoretical error for 217 newly measured masses is 0.737 MeV,
corresponding to an increase of 15\%.
This model therefore extrapolates to new regions somewhat less well than
does either the FRDM (1992)
or the ETFSI-1 (1992) model.
The deviations between
experimental and calculated masses for
these 217 new nuclei are shown in Fig.~\ref{devmtf}.
 
In the Wapstra-Audi mass
evaluations$\,^{2-4,49})$
there usually are listed not only
measured masses, but also masses estimated from systematic trends.
It is unfortunate that some theoretical mass studies have not observed the
difference between measured masses and masses given by systematic trends,
and included also the latter in the data set to which the model parameters
were adjusted. The masses given by systematic trends have to be considered
on the same basis as masses given by other models. We may therefore
study the reliability of the Wapstra-Audi systematic-trend model by
use of the methods applied above to other models. We consider the
masses given by systematic trends in the 1989 midstream evaluation$\,^{2})$
as model masses and compare them to the new masses determined in
the 1993 mass evaluation.$\,^{3,4})$\ess
The result is shown in Fig.\ \ref{devmsyst}. The systematic-trend model
error in the new region is 0.404 MeV, which is smaller than
the error of the FRDM (1992). However, the systematic-trend model
does not provide masses very far from known nuclei, and indeed
many of the newly measured masses were not predicted by the 1989 systematic-trend
model. Also, in this case it is not possible to provide
a ratio between the error for nuclei in the new region and
the error in the known region.
 
The discovery in late 1994 of the new element $Z=111$$\,^{30})$
allows us to test model extrapolateabilities in the region of large
proton numbers. In Fig.~\ref{comp3mod} we compare the experimental
results for the $\alpha$-decay chain of the heaviest known element
$^{272}$111 with predictions obtained in the FRDM (1992), the ETFSI-1
(1992) model$\,^{47,48})$, and
the 1992 fermion dynamical symmetry model of
Han, Wu, Feng, and Guidry.$\,^{54})$\ess  Clearly the predictions of the
FRDM (1992) agree much better with the new
experimental data than do those of the other two
models. The FRDM (1992) error for this chain is in fact considerably smaller than
the error in the region where the model constants were adjusted, whereas the
errors for the other two models, especially  the FDSM (1992),
are considerably larger than in the regions where
the constants of these two models were adjusted.
Because the FRDM (1992) $Q_{\alpha}$ error is
about 0.5 MeV for heavy nuclei in the known region,
one should not consistently
expect the exceptionally good agreement present in
Fig.~\ref{comp3mod} for all new
$\alpha$-decay chains that are discovered in the heavy region. However, the
agreement between experiment and predictions of
the FRDM (1992) seen in this figure, plus  better-than-expected agreement between
experiment and predictions of
the FRDM (1992) and the decay chains of another recently discovered element
with proton number $Z=110$,$\,^{29})$
confirm the conclusion reached earlier by
simulation$\,^{1})$ that the FRDM (1992) is reliable
to its stated accuracy as we move away from the
region of known elements towards the
superheavy region.

\section{ASTROPHYSICAL APPLICATIONS \label{astro}}

Nuclear physics and astrophysics share  several common themes:
(1) the nuclear reactions that are responsible
for nucleosynthesis and isotopic abundance patterns in
nature, (2) the energy sources of stellar events via static or
explosive nuclear burning, and (3) the behavior of nuclear matter at
and beyond nuclear densities, including
the equation of state for type-II supernova (SN II) explosions and
neutron stars.      
 
Here, we focus on explosive conditions in which large numbers of free
protons or neutrons are available. They can lead to the synthesis of
nuclei far from stability either via rapid proton capture and $\beta^+$
decay ($rp$-process) in novae and $X$-ray bursts or rapid neutron
capture and $\beta^-$ decay ($r$-process) in SN II\@. In both cases,
theoretical studies require  reaction rates and other nuclear
properties of unstable nuclei which, to a large extent, are not known
experimentally. Hence, a general understanding of their
nuclear-structure properties can  be obtained only through theoretical
means.              
 
Because several different nuclear quantities
are needed in $rp$- and $r$-process calculations, in the past it was
not possible to obtain them all from one source. Taking them from
different sources, however, raises questions of consistency.  In
such mixed-model calculations,$\,^{55})$\mss although
often performed due to the lack
of a unified approach, occasionally nuclear
structure signatures may vanish or artificial effects may occur, thus
strongly limiting their predictive power far from stability. Therefore,
attempts to use this approach to identify
the sites of  $rp$- and $r$-processes
may lead to erroneous conclusions.
Consequently, it is of great importance to provide
nuclear-structure properties based on a single, unified theoretical
framework within which all quantities of interest can be obtained.
We illustrate  this principle below with an example.
 
\subsection{The $rp$-process \label{rpproc}}
 
In a recent paper van Wormer {\it et al.\/}$\,^{56})$ have evaluated in
detail the nuclear-reaction sequences for conditions in explosive
hydrogen burning at temperatures beyond $10^8$ K,
corresponding to the
$rp$-process, a sequence of rapid proton-capture reactions and $\beta^+$
decays passing through proton-rich nuclei. Such processes typically
occur when hydrogen fuel is ignited under highly degenerate conditions
in explosive events on the surface of compact objects like white
dwarfs (novae) and neutron stars ($X$-ray bursts). Hydrogen
burning at high temperatures such as $2\times10^9$ K may also
occur in a stable fashion in so-called Thorne-Zytkow objects.
Such objects are expected to result from the merging of a neutron star
and a main-sequence star in a binary system. An $rp$-process
occurs then at the base of a fully convective envelope just above the
neutron core, giving the object the appearance of a red supergiant.
 
As pointed out by some authors,$\,^{56,57})$\mss
the $rp$-process cannot be
explained in a simple and clean fashion in terms of a
(p,$\gamma)\rightleftharpoons(\gamma$,p)
equilibrium  in all isotopic
chains. Under such circumstances, the nuclear data needed for a
theoretical description would be limited to nuclear masses and
corresponding proton separation energies  and
$\beta^+$-decay half-lives. However, the Coulomb barriers
in charged-particle captures lead to a cycle pattern with capture and
decay time scales of similar size. As outlined in Fig.~\ref{rpprocfig},
the $rp$-process at low temperatures is dominated by two successive
proton captures, starting out from an e-e nucleus, a $\beta^+$ decay,
a further proton capture into an even-$Z$ nucleus, another $\beta^+$
decay, and a final (p,$\alpha$) reaction close to stability.
 
The progress of the $rp$-process towards heavier nuclei depends
on the leakage ratio (p,$\gamma)/$(p,$\alpha$) into the next cycle.
Increasing temperature makes it possible to
overcome Coulomb barriers and extend
the cycles to more proton-rich nuclei,
which permits additional leakage via proton captures competing
with long $\beta$ decays. Beyond $3\times10^8$ K, all cycles break
open and a complete $rp$-pattern of proton
captures and $\beta$ decays is established, which may reach beyond
$^{56}$Ni.$\,^{56,57})$\ess

As a consequence, along with the separation energies and half-lives
that are required for an
equilibrium assumption, proton and $\alpha$-capture rates for
unstable nuclei are also required as nuclear-physics input into the
reaction network.   
In current calculations the majority of these reaction rates are not based
on measured cross sections, but are usually approximated by
simple statistical-model calculations.$\,^{58})$\ess
Even such statistical approaches require a number of input
quantities, such as 
(1) ground-state spins and parities of target, compound, and final
nuclei, (2) reaction energy releases, (3) realistic
nucleon-nucleus and $\alpha$-nucleus optical potentials,
(4) giant-dipole resonance energies and widths, and
(5) level-density models.
It is desirable to improve the accuracy of current level-density
models by utilizing shell and pairing corrections of modern mass models.
 
\subsection{The $r$-process}
 
The $r$-process was inferred$\,^{59,60})$ from the
observation of characteristic peaks in the abundance curve of
$\beta$-stable nuclei. The peak maxima are located 4, 4, and 9
neutrons {\it below} the magic neutron numbers 50, 82, and 126,
respectively. In Fig.~\ref{rproc} some basic features of the
$r$-process are outlined.  The position of the $r$-process line depends
on nuclear-structure properties and the stellar conditions under which
it occurs, in particular the temperature, density, and duration of
the neutron flux.  The line plotted in the figure corresponds to a
neutron separation energy of 2.4 MeV\@. The $r$-process is dynamic and
may be located anywhere in the region 1.5 MeV $<S_{\rm 1n}<3.0$ MeV\@,
depending on the stellar conditions.  Magic neutron numbers play a
special role in the $r$-process as is partially seen in
Fig.~\ref{rproc}, where the $r$-process line has a kink at each magic
neutron number.  In addition, nuclei pile up at these magic
neutron numbers because of the long $\beta$-decay half-lives  and
sudden lowering  of the neutron-capture cross section that occurs at
magic neutron numbers.  After freezeout  of the neutron flux, these
nuclei  $\beta$ decay back towards the line of $\beta$ stability.  It
is easy to see from the figure that the peaks in the abundance curve
are related to a concentration of neutron-magic nuclei {\it far from
$\beta$ stability}.  In fact, this effect has been used to put rough
constraints on nuclear masses far from $\beta$ stability and to rule
out nuclear mass models with predictions far from these  constraints.
The decay back to the line of $\beta$ stability is also influenced by
$\beta$-delayed neutron emission.
 
Since the pioneering work of Burbridge {\it et al.\/}$\,^{59})$
and Cameron$\,^{60})$ the rapid neutron capture process has been
associated with explosive environments with high temperature $T\geq10^9$ K
and high neutron-number density
$n_{\rm n}\geq 10^{20}$~cm$^{-3}$. Under such conditions, the neutron-capture
time scales of heavy nuclei are so short that within about
$10^{-4}$ s highly neutron-rich nuclei can be produced up to 15
to 30 mass units away from $\beta$ stability with neutron separation
energies $S_{\rm 1n}\approx 1.5$ to 3.0 MeV\@. Neutron captures are not
hindered by increasing Coulomb barriers, in contrast to the case for the
$rp$-process in the previous section. Magic neutron numbers are
encountered for smaller mass numbers $A\/$ than in the valley of
stability, which, after freezeout  and $\beta$ decay, shifts the
observed solar-system $r$-process abundance peaks below the $s$-process
peaks. However,  besides this basic understanding, the history of $r$-process
research has been quite diverse in suggested astrophysical scenarios,
as well as with  the required size of the nuclear network.$\,^{55})$
 
The observed isotopic $r$-abundances $N_{r,\odot}$ shown in
Fig.~\ref{rproc} are the result of successive neutron captures along
the $r$-process path and $\beta^-$ decay back to stability, thus
depending, apart from stellar parameters, on a variety of nuclear
properties of nuclei with extreme $N/Z\/$ ratios.  In the general case,
among nuclear-physics quantities, ground-state masses and corresponding
$Q_{\beta}$ values, neutron-separation energies $S_{\rm 1n}$,
$\beta^-$-decay half-lives, probabilities of $\beta$-delayed neutron emission
$P_{\rm n}$, neutron-capture cross sections, and ground-state spins and
deformations are of importance in $r$-process
calculations.$\,^{55,61})$\ess For the first time, most of
these quantities can be obtained from a single, unified model, namely our
present macroscopic-microscopic model.  Neutron-capture rates are so
far calculated with the statistical Hauser-Feshbach model, as long as
the level density in the compound nucleus is sufficiently high to
justify such an approach.  As is the case for proton captures discussed
in Sec.~\ref{rpproc}, this method requires a knowledge of additional
quantities, such as optical potentials and giant-dipole-resonance
and level-density parameters. For nuclei in the $r$-process path, in
particular near closed neutron shells, the level density is small and
the Hauser-Feshbach approach might no longer be applicable. In these
cases, Breit-Wigner resonance capture and direct capture have to be
considered, again requiring additional nuclear-physics properties as
input for the respective model
calculations.$\,^{61,62})$\ess For nuclei with $Z\geq80$,
fission barriers and rates of $\beta$-delayed as well as
neutron-induced fission are also important.  Finally, within the
recently favored ``hot-entropy-bubble'' $r$-process
scenario,$\,^{63,64})$\mss charged-particle
reactions during the so-called $\alpha$-rich freezeout are also required in
the $A\approx80$ region. Taken together, all these nuclear data for
thousands of mostly unknown isotopes require a huge reaction network
for ``complete'' $r$-process calculations.
 
Primarily in order to facilitate these complicated and time-consuming
calculations, since 1957 many attempts to predict the $N_{r,\odot}$
distribution were based on the simplified assumption of the
(n,$\gamma)\rightleftharpoons (\gamma$,n) equilibrium
concept.$\,^{55,59,61})$\ess When assuming in
addition a steady-flow equilibrium of $\beta$ decays, the prediction of
$r$-abundances requires only the input of nuclear masses and
corresponding neutron separation energies $S_{\rm 1n}$, $\beta$-decay
half-lives $T_{\beta}$ and $\beta$-delayed neutron-emission
probabilities $P_{\rm n}$, as well as the stellar parameters $T_{9}$,
$n_{\rm n}$, and  the process duration. Whereas for a given $n_{\rm n}$
the $S_{\rm 1n}$ determine the $r$-process path, the $T_{\beta}$ of the
isotopes along this flow path determine, in principle, the progenitor
abundances and, when $P_{\rm n}$ branching during freezeout is
considered, also the final $r$-abundances. Only in recent years could
the validity of this ``waiting-point'' approximation in combination
with a steady $\beta$-decay flow be confirmed locally for the $A\approx
80$ and 130 $N_{r,\odot}$ peaks on the basis of the first experimental
information in the $r$-process path.$\,^{65})$\ess These recent
results showed clearly, for example,  how the long $T_{\beta}$ of the
classical $N=82$ waiting-point nucleus $^{130}$Cd  directly correlated
with the large $N_{r,\odot}$ value of its isobar $^{130}$Te in the
$A\approx 130$ abundance peak seen in Fig.~\ref{rproc}.
 
Since there is not yet complete consensus on the stellar site of the
$r$-process and the specific astrophysical conditions under which it
takes place, one deductive approach to theoretical $r$-process studies
has been to take the $N_{r,\odot}$ observables$\,^{66})$ as a
constraint that allow the derivation of the necessary conditions
required to reproduce these features. Using the unified nuclear-physics
basis presented in this paper, supplemented by all experimental data
available up to 1991 as well as local improvements of the QRPA
calculations, Kratz {\it et al.\/}$\,^{61})$ have obtained
significant progress in reproducing the $N_{r,\odot}$ pattern relative
to the situation five or ten years ago. As an example, we show in
Fig.~\ref{rabu1} results of $r$-process calculations for two different
mass models. The top calculation  is based on our 1991
version of the FRDM.$\,^{45,67})$\ess
Our current FRDM (1992)  shown in
Fig.~\ref{masd92a} has smaller mass and $S_{\rm 1n}$ errors, but our
preliminary $r$-process studies with it do not show significantly improved
agreement. In the lower part of Fig.~\ref{rabu1}, nuclear masses and
input for the QRPA calculations were taken from the ETFSI-1 (1992) model
in its preliminary, privately circulated version.
 
With the current agreement between  $N_{r,\odot}$ data and the
calculations, the theoretical treatment is sufficiently accurate
that some conclusions can be drawn about the stellar conditions
responsible for the production of $r$-process nuclei. For example, one
has found that it is not possible to reproduce the $N_{r,\odot}$ curve
assuming a global steady-flow process. Instead, a minimum of three
$r$-components with different neutron densities is required. Each of
the components proceeds up to one of the abundance peaks and reaches a
local steady-flow equilibrium which breaks down at the top of each
peak, situated at one of the $N=50$, 82, and 126 magic shells. With these
results,  the explosive He-burning scenarios favored in the 1980s can
definitely be ruled out as possible sites for the $r$-process.
 
The large deviations from the $N_{r,\odot}$ pattern evident
in Fig.~\ref{rabu1}, especially those just before the
abundance peaks at $A\approx 130$ and 195, were
interpreted$\,^{57,68,69})$ as arising from
overly strong magic-neutron shell corrections as one moves away from a
doubly magic configuration, from the neglect of the
proton-neutron residual interaction, and from correlated problems with
describing shape transitions in the neutron mid-shell regions around
$N=66$ and 104.     
 
Rather than representing a failure of the FRDM (1992)
far from stability, these difficulties are due to
normal inaccuracies that occur anywhere in the chart of the nuclides,
both near stability and far from stability. They must be expected and
considered normal in a model based on such a simple
effective interaction as the one-body
single-particle potential with a simple pairing residual interaction.
Despite some deficiencies, the current
models, when used appropriately,
have been {\it sufficiently accurate\/} to {\it considerably
advance} our understanding of several astrophysical processes
and to {\it identify} specific nuclear-structure features
of nuclei far from stability near magic shells and close to
the neutron drip lines. These nuclei are normally inaccessible
to experiment, but unique signatures of their nuclear
structure are well preserved as differences in the observed
$N_{r,\odot}$ pattern relative to the pattern calculated with
the FRDM (1992)\@. Our identification of specific nuclear-structure
features in nuclei far from stability has generated substantial interest
in the nuclear-physics community and stimulated several
calculations of nuclear masses in limited regions far from stability
in terms of self-consistent mean-field
theories.$\,^{70-72})$\ess
Some of these local calculations now produce
nuclear masses that remove some of the discrepancies that are
present in Fig.~\ref{rabu1}.
 
However, to be considered an improvement over current global
nuclear-structure calculations, a new theory must achieve more than
just a better local description of  features that were postulated on
the basis of existing theories. To justify the designation new and
improved,  a new theory should globally reproduce better nuclear masses
and other ground-state nuclear-structure effects than does the FRDM (1992)\@.
It should achieve these results with a simple, global choice of
constants.  And, it should, just as has the current FRDM (1992)\@, correctly
predict new nuclear-structure features far from stability and serve to
further enhance our understanding of several astrophysical processes.
The results achieved so far in new mean-field calculations are
important preliminary steps in this direction.

We have seen that certain differences in the abundance fits and remaining
deficiencies can be attributed to the nuclear mass models applied. We now want
to check whether effects from different half-life sets can also be separated
out in our $N_{r,\odot}$ calculations.
 
As we have discussed earlier,$\,^{61})$\mss once the $r$-process
path is defined by a contour line of constant $S_{\rm 1n}$ values,
the abundances of isotopes in this path are directly related to their
$\beta$-decay half-lives according to the waiting-point
concept.$\,^{59})$\ess
These $\beta$-decay half-lives are determined by the energy
window $Q_{\beta}$, 
on the one hand, and by the low-lying nuclear structures in the Gamow-Teller
strength function, on the other hand. As has been discussed in detail
elsewhere,$\,^{61})$\mss $T_{\beta}$ is predominantly influenced by the
latter properties. Therefore, statistical models such as the gross
theory of $\beta$ decay,$\,^{73})$\ess
which neglect nuclear structures,  will not be
able to describe the $\beta$-decay quantities far from stability in an
adequate way. It has been well known for more than a decade that, due to the
missing nuclear-structure effects, the ${\beta}$-decay half-lives
from this model are 
systematically too long by factors of five to 10 far from stability.
Nevertheless,  because the gross-theory
predictions are still occasionally used in astrophysical
calculations,$\,^{64,74})$\mss we have for the purpose
of illustrating its deficiencies performed some $T_{\beta}$ and
$P_{\rm n}$ calculations with this model using $Q_{\beta}$ values
from both the FRDM (1992) and ETFSI-1 (1992) model. We can show
that the  differences in the $Q_{\beta}$ predictions of the two models
for isotopes in the regions of the  $r$-process path affect the $T_{\beta}$
values obtained in the gross theory only to a small extent. Therefore
the $T_{\beta}$ effects in the calculated
$r$-abundances originate mainly from the missing nuclear structures in
the $\beta$-strength functions of the gross theory.
 
In Fig.~\ref{rabu2} we present a comparison of $r$-process calculations
based on the gross theory and on the QRPA\@. The upper part of
Fig.~\ref{rabu2} shows that for our best-fit conditions of the third
component,$\,^{61})$\mss a switch from the $T_{\beta}$(QRPA)
consistent with the FRDM masses to $T_{\beta}$(g.t.) creates
$r$-overabundances in the range $150\leq A\leq 180$. It is possible to
adjust for this artificial effect by a change of astrophysical
conditions, for example, by increasing $n_{\rm n}$ in order to obtain a
more neutron-rich $r$-process path and shorter $T_{\beta}$(g.t.) in
that mass region.$\,^{74})$\ess However, as can be seen in the
lower left-hand part of Fig.~\ref{rabu2}, higher neutron densities, which
improve the $N_{r,\odot}$ fits in the $150\leq A\leq 180$ range, but
are difficult to obtain in realistic $r$-process
scenarios,$\,^{63,64})$\mss shift the structure of
the $A\approx 195$ peak too far to lower masses and make such an
approach invalid. We can only conclude that far from $\beta$ stability
the values of $T_{\beta}$(g.t.) are, in fact, too long.

\section{SINGLE-PARTICLE LEVELS \label{singlev}}
 
As a final result we present in
Figs.~\ref{p008016}--\ref{n124308}
calculated proton and neutron single-particle
level diagrams for representative spherical and
deformed nuclei throughout the periodic system.
The diagrams are useful for obtaining the spin and
parity of low-lying states of odd-even nuclei and for
identifying gaps in the level spectra that may be associated
with particularly stable proton-neutron combinations.
 
For heavier nuclei higher-multipole deformations become increasingly
important, as can be seen in the mass table and color overview figures
in our mass calculations.$\,^{1})$. For prolate shapes we have
therefore chosen $\epsilon_4$ and $\epsilon_6$ to be functions of
$\epsilon_2$. The hexadecapole deformation $\epsilon_4$ is either a linear
function of $\epsilon_2$ in the entire range $0\leq \epsilon_2\leq 0.4$ or
linear in the range  $0\leq \epsilon_2\leq 0.2$ and constant in the range
$\epsilon_2\geq 0.2$, as indicated on the upper horizontal axis.
The  hexacontratetrapole deformation $\epsilon_6$
is zero when it is not explicitly mentioned in the figure captions.
Otherwise, it varies
linearly in the entire range $0\leq \epsilon_2\leq 0.4$
in the manner       
indicated in the figure captions.
 
One notes that proton number $Z=14$ and neutron number $N=14$
correspond to well-developed spherical gaps for several of the lighter
systems. The effect of the higher-multipole deformation parameters
$\epsilon_4$ and $\epsilon_6$ is clearly visible
in the level-diagram sequences shown in Figs.~\ref{p040080}--\ref{n040106}
and \ref{p100240}--\ref{n100262}, for example.
 
In the heavy region it is particularly interesting that the
deformed shell gaps at proton numbers $Z=104$--110 and neutron number
$N=162$ emerge only for relatively  large positive values of the
hexadecapole deformation parameter $\epsilon_4$. These gaps give rise
to unusual stability, which has made possible the discovery of several
new elements in this
region.$\,^{26-30})$\ess
 
Figure \ref{n114298} shows the  predicted large spherical neutron
gap $N=184$. However, the unusually large density of
single-particle levels just above has the consequence that the
largest negative microscopic correction occurs approximately at
neutron number $N=178$ in our model.\\[3ex]
{\it Acknowledgements}\\
This work was supported by the U.\ S.\ Department of Energy
and the German Ministry for Research and Technology (BMFT grant 06MZ465).
We are grateful to  
A.\ Iwamoto, J.\ M.\ Pearson, H. Sagawa, and F.-K.\ Thielemann for many
useful discussions. The initial versions of the codes used to
calculate $\beta$-decay properties were written in collaboration
with J.\ Krumlinde. 
 
\newpage            
\markboth           
{\it P. M\"{o}ller, J. R. Nix, and K.-L. Kratz/Nuclear
Properties}         
{\it P. M\"{o}ller, J. R. Nix, and K.-L. Kratz/Nuclear
Properties}         
\begin{center}                                                                  
{\bf References}                                                                
\end{center}                                                                    
\newcounter{bona}                                                               
\begin{list}%
{\arabic{bona})}{\usecounter{bona}                                              
\setlength{\leftmargin}{0.5in}                                                  
\setlength{\rightmargin}{0.0in}                                                 
\setlength{\labelwidth}{0.3in}                                                  
\setlength{\labelsep}{0.15in}                                                   
}                                                                               
\item                                                                           
P.\ M{\"{o}}ller, J.\ R.\ Nix, W.\ D.\ Myers, and W.\ J.\ Swiatecki, {Atomic    
  Data Nucl.\ Data Tables} {\bf 59} (1995) 185.                                 
                                                                                
\item                                                                           
G.\ Audi, Midstream atomic mass evaluation, private communication (1989), with  
  four revisions.                                                               
                                                                                
\item                                                                           
G.\ Audi and A.\ H.\ Wapstra, Nucl.\ Phys.\ {\bf A565} (1993) 1.                
                                                                                
\item                                                                           
G.\ Audi and A.\ H.\ Wapstra, Nucl.\ Phys.\ {\bf A565} (1993) 66.               
                                                                                
\item                                                                           
V.\ E.\ Viola, Jr.\ and G.\ T.\ Seaborg, J.\ Inorg.\ Nucl.\ Chem.\ {\bf 28}     
  (1966) 741.                                                                   
                                                                                
\item                                                                           
A.\ Sobiczewski, Z.\ Patyk, and S.\ \v{C}wiok, Phys.\ Lett.\ {\bf B224} (1989)  
  1.                                                                            
                                                                                
\item                                                                           
P.\ M{\"{o}}ller and J.\ R.\ Nix, Nucl.\ Phys.\ {\bf A536} (1992) 20.           
                                                                                
\item                                                                           
{\AA}.\ Bohr, B.\ R.\ Mottelson, and D.\ Pines, Phys.\ Rev.\ {\bf 110} (1958)   
  936.                                                                          
                                                                                
\item                                                                           
S.\ T.\ Belyaev, Kgl.\ Danske Videnskab.\ Selskab.\ Mat.-Fys.\ Medd.\ {\bf      
  31}:No.\ 11 (1959).                                                           
                                                                                
\item                                                                           
S.\ G.\ Nilsson and O.\ Prior, Kgl.\ Danske Videnskab.\ Selskab.\ Mat.-Fys.\    
  Medd.\ {\bf 32}:No.\ 16 (1961).                                               
                                                                                
\item                                                                           
W.\ Ogle, S.\ Wahlborn, R.\ Piepenbring, and S.\ Fredriksson, Rev.\ Mod.\       
  Phys.\ {\bf 43} (1971) 424.                                                   
                                                                                
\item                                                                           
H.\ J.\ Lipkin, Ann.\ Phys.\ (N.\ Y.) {\bf 9} (1960) 272.                       
                                                                                
\item                                                                           
Y.\ Nogami, Phys.\ Rev.\ {\bf 134} (1964) B313.                                 
                                                                                
\item                                                                           
H.\ C.\ Pradhan, Y.\ Nogami, and J.\ Law, Nucl.\ Phys.\ {\bf A201} (1973) 357.  
                                                                                
\item                                                                           
E.\ R.\ Cohen and B.\ N.\ Taylor, CODATA Bull.\ No.\ {\bf 63} (1986).           
                                                                                
\item                                                                           
E.\ R.\ Cohen and B.\ N.\ Taylor, Rev.\ Mod.\ Phys.\ {\bf 59} (1987) 1121.      
                                                                                
\item                                                                           
J.\ Krumlinde and P.\ M{\"{o}ller}, Nucl.\ Phys.\ {\bf A417} (1984) 419.        
                                                                                
\item                                                                           
P.\ M{\"{o}}ller and J.\ Randrup, Nucl.\ Phys.\ {\bf A514} (1990) 1.            
                                                                                
\item                                                                           
I.\ Hamamoto, Nucl.\ Phys.\ {\bf 62} (1965) 49.                                 
                                                                                
\item                                                                           
J.\ A.\ Halbleib, Sr.\ and R.\ A.\ Sorensen, Nucl.\ Phys.\ {\bf A98} (1967)     
  542.                                                                          
                                                                                
\item                                                                           
J.\ Engel, P.\ Vogel, and M.\ R.\ Zirnbauer, Phys.\ Rev.\ {\bf C37} (1988) 731. 
                                                                                
\item                                                                           
B.\ Lauritzen, Nucl.\ Phys.\ {\bf A489} (1988) 237.                             
                                                                                
\item                                                                           
A.\ deShalit and H.\ Feshbach, Theoretical physics, vol.\ I: Nuclear structure  
  (Wiley, New York, 1974).                                                      
                                                                                
\item                                                                           
M.\ A.\ Preston, Physics of the nucleus (Addison-Wesley, Reading, 1962).        
                                                                                
\item                                                                           
N.\ B.\ Gove and M.\ J.\ Martin, {Nucl.\ Data Tables} {\bf 10} (1971) 205.      
                                                                                
\item                                                                           
G.\ {M\"{u}nzenberg}, S.\ Hofmann, F.\ P.\ He{\ss}berger, W.\ Reisdorf, K.-H.\  
  Schmidt, J.~R.~H.\ Schneider, P.\ Armbruster, C.-C.\ Sahm, and B.\ Thuma, Z.\ 
  Phys.\ {\bf A300} (1981)~7.                                                   
                                                                                
\item                                                                           
G.\ {M\"{u}nzenberg}, P.\ Armbruster, F.\ P.\ He{\ss}berger, S.\ Hofmann, K.\   
  Poppensieker, W.\ Reisdorf, J.\ R.\ H.\ Schneider, W.\ F.\ W.\ Schneider,     
  K.-H.\ Schmidt, C.-C.\ Sahm, and D.\ Vermeulen, Z.\ Phys.\ {\bf A309} (1982)  
  89.                                                                           
                                                                                
\item                                                                           
G.\ {M\"{u}nzenberg}, P.\ Armbruster, H.\ Folger, F.\ P. He{\ss}berger, S.\     
  Hofmann, J.\ Keller, K.\ Poppensieker, W.\ Reisdorf, K.-H.\ Schmidt, H.\ J.\  
  {Sch\"{o}tt}, M.\ E.\ Leino, and R.\ Hingmann, Z.\ Phys.\ {\bf A317} (1984)   
  235.                                                                          
                                                                                
\item                                                                           
S.\ Hofmann, N.\ Ninov, F.\ P.\ He{\ss}berger, P.\ Armbruster, H.\ Folger, G.\  
  {M\"{u}nzenberg}, H.\ J.\ Sch{\"{o}}tt, A.\ G.\ Popeko, A.\ V.\ Yeremin, A.\  
  N.\ Andreyev, S.\ Saro, R.\ Janik, and M.\ Leino, Z.\ Phys.\ {\bf A350}       
  (1995) 277.                                                                   
                                                                                
\item                                                                           
S.\ Hofmann, N.\ Ninov, F.\ P.\ He{\ss}berger, P.\ Armbruster, H.\ Folger, G.\  
  {M\"{u}nzenberg}, H.\ J.\ Sch{\"{o}}tt, A.\ G.\ Popeko, A.\ V.\ Yeremin, A.\  
  N.\ Andreyev, S.\ Saro, R.\ Janik, and M.\ Leino, Z.\ Phys.\ {\bf A350}       
  (1995) 281.                                                                   
                                                                                
\item                                                                           
E.\ O.\ Fiset and J.\ R.\ Nix, Nucl.\ Phys.\ {\bf A193} (1972) 647.             
                                                                                
\item                                                                           
P.\ {M\"{o}ller}, S.\ G.\ Nilsson, and J.\ R.\ Nix, Nucl.\ Phys.\ {\bf A229}    
  (1974) 292.                                                                   
                                                                                
\item                                                                           
P.\ {M\"{o}ller} and J.\ R.\ Nix, Nucl.\ Phys.\ {\bf A361} (1981) 117.          
                                                                                
\item                                                                           
D.\ G.\ Madland and J.\ R.\ Nix, Nucl.\ Phys.\ {\bf A476} (1988) 1.             
                                                                                
\item                                                                           
K.-L.\ Kratz, H.\ Ohm, A.\ Schr{\"{o}}der, H.\ Gabelmann, W.\ Ziegert, H.\ V.\  
  Klapdor, J.\ Metzinger, T.\ Oda, P.\ Pfeiffer, G.\ Jung, L.\ Alquist, and G.\ 
  I.\ Crawford, Proc.\ 4th Int.\ Conf.\ on nuclei far from stability,           
  Helsing{\o}r, 1981, Report No.\ CERN-81-09 (1981) p.\ 317.                    
                                                                                
\item                                                                           
K.-L.\ Kratz, Nucl.\ Phys.\ {\bf A417} (1984) 447.                              
                                                                                
\item                                                                           
R.\ Bengtsson, P.\ {M\"{o}ller}, J.\ R.\ Nix, and Jing-ye Zhang, Phys.\ Scr.\   
  {\bf 29} (1984) 402.                                                          
                                                                                
\item                                                                           
E.\ Browne, private communication (1988).                                       
                                                                                
\item                                                                           
E.\ Bender, K.\ Muto, and H.\ V.\ Klapdor, Phys.\ Lett.\ {\bf B208} (1988) 53.  
                                                                                
\item                                                                           
J.\ M.\ Nitschke, P.\ A.\ Wilmarth, J.\ Gilat, P.\ M{\"{o}}ller, and K.\ S.\    
  Toth, Proc.\ Fifth.\ Int.\ Conf.\ on nuclei far from stability, Rosseau Lake, 
  Ontario, Canada, 1987, AIP Conf.\ Proc.\ {\bf 164} (AIP, New York, 1988) p.\  
  403.                                                                          
                                                                                
\item                                                                           
J.\ M.\ Nitschke, P.\ A.\ Wilmarth, R.\ B.\ Firestone, P.\ M{\"{o}}ller, K.\    
  S.\ Toth, and J.\ Gilat, Phys.\ Rev.\ Lett.\ {\bf 62} (1988) 2805.            
                                                                                
\item                                                                           
G.\ {M\"{u}nzenberg}, P.\ Armbruster, S.\ Hofmann, F.\ P.\ He{\ss}berger, H.\   
  Folger, J.\ G.\ Keller, V.\ Ninov, K.\ Poppensieker, A.\ B.\ Quint, W.\       
  Reisdorf, K.-H.\ Schmidt, J.\ R.\ H.\ Schneider, H.\ J.\ Sch{\"{o}tt}, K.\    
  S{\"{u}}mmerer, I.\ Zychor, M.\ E.\ Leino, D.\ Ackermann, U.\ Gollerthan, E.\ 
  Hanelt, W.\ Morawek, D.\ Vermeulen, Y.\ Fujita, and T.\ Schwab, Z.\ Phys.\    
  {\bf A333} (1989) 163.                                                        
                                                                                
\item                                                                           
P.\ M{\"{o}}ller and J.\ R.\ Nix, {Atomic Data Nucl.\ Data Tables} {\bf 39}     
  (1988) 213.                                                                   
                                                                                
\item                                                                           
P.\ M{\"{o}}ller and J.\ R.\ Nix, Nucl.\ Phys.\ {\bf A549} (1992) 84.           
                                                                                
\item                                                                           
P.\ {M\"{o}ller} and J.\ R.\ Nix, Proc.\ 6th Int.\ Conf.\ on nuclei far from    
  stability and 9th Int.\ Conf.\ on nuclear masses and fundamental constants,   
  Bernkastel-Kues, 1992 (IOP Publishing, Bristol, 1993) p.\ 43.                 
                                                                                
\item                                                                           
P.\ {M\"{o}ller}, J.\ R.\ Nix, K.-L.\ Kratz, A.\ W{\"{o}}hr, and F.-K.\         
  Thielemann, Proc.\ First Symp.\ on nuclear physics in the universe, Oak       
  Ridge, 1992 (IOP Publishing, Bristol, 1993) p.\ 433.                          
                                                                                
\item                                                                           
Y.\ Aboussir, J.\ M.\ Pearson, A.\ K.\ Dutta, and F.\ Tondeur, Nucl.\ Phys.\    
  {\bf A549} (1992) 155.                                                        
                                                                                
\item                                                                           
Y.\ Aboussir, J.\ M.\ Pearson, A.\ K.\ Dutta, and F.\ Tondeur, Atomic Data      
  Nucl.\ Data Tables {\bf 61} (1995) 127.                                       
                                                                                
\item                                                                           
A.\ H.\ Wapstra, G.\ Audi, and R.\ Hoekstra, {Atomic Data Nucl.\ Data Tables}   
  {\bf 39} (1988) 281.                                                          
                                                                                
\item                                                                           
H.\ von Groote, E.\ R.\ Hilf, and K.\ Takahashi, Atomic Data Nucl.\ Data Tables 
  {\bf 17} (1976) 418.                                                          
                                                                                
\item                                                                           
A.\ H.\ Wapstra and K.\ Bos, {Atomic Data Nucl.\ Data Tables} {\bf 19} (1977)   
  175.                                                                          
                                                                                
\item                                                                           
W.\ D.\ Myers and W.\ J.\ Swiatecki, Lawrence Berkeley Laboratory Report No.\   
  LBL-36803 (1994).                                                             
                                                                                
\item                                                                           
W.\ D.\ Myers and W.\ J.\ Swiatecki, Nucl.\ Phys.\ {\bf A}, to be published.    
                                                                                
\item                                                                           
X.-L.\ Han, C.-L.\ Wu, D.\ H.\ Feng, and M.\ W.\ Guidry, Phys.\ Rev.\ {\bf C45} 
  (1992) 1127.                                                                  
                                                                                
\item                                                                           
J.\ J.\ Cowan, F.-K.\ Thielemann, and J.\ W.\ Truran, Phys.\ Rep.\ {\bf 208}    
  (1991) 267.                                                                   
                                                                                
\item                                                                           
L.\ Van Wormer, J.\ G{\"{o}}rres, C.\ Iliadis, M.\ Wiescher, and F.-K.\         
  Thielemann, Ap.\ J.\ {\bf 432} (1994) 326.                                    
                                                                                
\item                                                                           
F.-K.\ Thielemann, K.-L.\ Kratz, B.\ Pfeiffer, T.\ Rauscher, L.\ Van Wormer,    
  and M.\ C.\ Wiescher, Nucl.\ Phys.\ {\bf A570} (1994) 329c.                   
                                                                                
\item                                                                           
F.-K.\ Thielemann, M.\ Arnould, and J.\ W.\ Truran, Advances in nuclear         
  astrophysics, (Les Editions Fronti{\`e}res, Gif sur Yvette, 1987) p.\ 525.    
                                                                                
\item                                                                           
E.\ M.\ Burbridge, G.\ R.\ Burbridge, A.\ A.\ Fowler, and F.\ Hoyle, Rev.\      
  Mod.\ Phys.\ {\bf 29} (1957) 547.                                             
                                                                                
\item                                                                           
A.\ G.\ W.\ Cameron, Atomic Energy of Canada Report No.\ CRL-41 (1957).         
                                                                                
\item                                                                           
K.-L.\ Kratz, J.-P.\ Bitouzet, F.-K.\ Thielemann, P.\ M{\"{o}}ller, and B.\     
  Pfeiffer, Ap.\ J.\ {\bf 403} (1993) 216.                                      
                                                                                
\item                                                                           
A.\ W{\"{o}}hr, W.\ B{\"{o}}hmer, S.\ Schoedder, K.-L.\ Kratz, H.\ Huber, E.\   
  Krausmann, H.\ Krauss, H.\ Oberhummer, T.\ Rauscher, and F.-K.\ Thielemann,   
  Proc.\ 8th Int.\ Symp.\ on capture gamma-ray spectroscopy and related topics, 
  Fribourg, 1993 (World Scientific, Singapore, 1994) p.\ 762.                   
                                                                                
\item                                                                           
S.\ E.\ Woosley and R.\ Hoffman, Ap.\ J.\ {\bf 395} (1992) 202.                 
                                                                                
\item                                                                           
K.\ Takahashi, J.\ Witti, and H.-T.\ Janka, Astron.\ Astrophys.\ {\bf 286}      
  (1994) 857.                                                                   
                                                                                
\item                                                                           
K.-L.\ Kratz, Rev.\ Mod.\ Astron.\ {\bf 1} (1988) 184.                          
                                                                                
\item                                                                           
F.\ K{\"{a}}ppeler, H.\ Beer, and K.\ Wisshak, Rep.\ Prog.\ Phys.\ {\bf 52}     
  (1989) 945.                                                                   
                                                                                
\item                                                                           
P.\ M{\"{o}}ller, J.\ R.\ Nix, W.\ D.\ Myers, and W.\ J.\ Swiatecki, Nucl.\     
  Phys.\ {\bf A536} (1992) 61.                                                  
                                                                                
\item                                                                           
K.-L.\ Kratz, P.\ M{\"{o}}ller, B.\ Pfeiffer, and F.-K.\ Thielemann, Proc.\ 8th 
  Int.\ Symp.\ on capture gamma-ray spectroscopy and related topics, Fribourg,  
  1993 (World Scientific, Singapore, 1994) p.\ 724.                             
                                                                                
\item                                                                           
K.-L.\ Kratz, Proc.\ 3rd Int.\ Symp.\ on nuclear astrophysics: ``nuclei in the  
  cosmos III,'' Gran Sasso, 1994, AIP Conference Proceedings No.\ 327 (American 
  Institute of Physics, New York, 1994) p.\ 113.                                
                                                                                
\item                                                                           
J.\ Dobaczewski, I.\ Hamamoto, W.\ Nazarewicz, and J.\ A.\ Sheikh, Phys.\ Rev.\ 
  Lett.\ {\bf 72} (1994) 381.                                                   
                                                                                
\item                                                                           
N.\ Fukunishi, T.\ Otsuka, and I.\ Tanihata, Phys.\ Rev.\ {\bf C48} (1993)      
  1648.                                                                         
                                                                                
\item                                                                           
B.\ Chen, J.\ Dobaczewski, K.-L.\ Kratz, K.\ Langanke, B.\ Pfeiffer, F.-K.\     
  Thielemann, and P.\ Vogel, Phys.\ Lett.\ {\bf B355} (1995) 37.                
                                                                                
\item                                                                           
K.\ Takahashi, M.\ Yamada, and T.\ Kondoh, {Atomic Data Nucl.\ Data Tables}     
  {\bf 12} (1973) 101.                                                          
                                                                                
\item                                                                           
W.\ M.\ Howard, S.\ Gorierly, M.\ Rayet, and M.\ Arnould, Ap.\ J.\ {\bf 467}    
  (1993) 713.                                                                   
                                                                                
\end{list}                                                                      
\markboth           
{\it P. M\"{o}ller, J. R. Nix, and K.-L. Kratz/Nuclear
Properties}         
{\it P. M\"{o}ller, J. R. Nix, and K.-L. Kratz/Nuclear
Properties}\mbox{ } \\ [2ex]
\newpage            
\begin{center}      
{\Large {\bf Figure captions}}\\[4ex]
\end{center}        
\begin{list}        
{\Roman{bean}}{\usecounter{bean}
\setlength{\leftmargin}{1.0in}
\setlength{\rightmargin}{0.0in}
\setlength{\labelwidth}{0.75in}
\setlength{\labelsep}{0.25in}
}                   
\item[Fig.\ \ref{lippdpl2}\hfill]
Global microscopic pairing gap for protons. The single-particle levels
entering the pairing calculation correspond to ground-state shapes that
have been determined by minimizing the total potential energy with
respect to $\epsilon_2$, $\epsilon_4$, $\epsilon_3$, and $\epsilon_6$
shape degrees of freedom.  The jagged black lines
indicate regions where experimental proton pairing
gaps may be extracted from fourth-order odd-even mass differences.  Magic proton and
neutron numbers are indicated by pairs of thin, parallel lines.  In the
Lipkin-Nogami model it is the sum $\Delta_{\rm p} +\lambda_{\rm 2p}$ plotted here
that should be compared to odd-even mass differences.  In contrast to
the behavior of BCS solutions, this  sum  shows no sign of collapse at
magic proton numbers.
 
\item[Fig.\ \ref{lipndpl2}\hfill]
Global microscopic pairing gap for neutrons. The single-particle levels
entering the pairing calculation correspond to ground-state shapes that
have been determined by minimizing the total potential energy with
respect to $\epsilon_2$, $\epsilon_4$, $\epsilon_3$, and $\epsilon_6$
shape degrees of freedom.  The jagged black lines
indicate regions where experimental neutron pairing
gaps may be extracted from fourth-order odd-even mass differences.  Magic proton and
neutron numbers are indicated by pairs of thin, parallel lines.  In the
Lipkin-Nogami model it is the sum $\Delta_{\rm n} +\lambda_{\rm 2n}$ plotted here
that should be compared to odd-even mass differences.  In contrast to
the behavior of BCS solutions, this sum shows no sign of collapse at
magic neutron numbers.
 
\item[Fig.\ \ref{errpln}\hfill]
Discrepancy between experimental proton pairing gaps determined from
fourth-order odd-even mass differences and microscopic pairing gaps $\Delta_{{\rm
LN}_{\rm p}}= \Delta_{\rm p} +\lambda_{\rm 2p}$ obtained in the Lipkin-Nogami
model.  Magic proton and neutron numbers are indicated by pairs of
thin, parallel lines.  Fairly large discrepancies occur in several
places. However, these discrepancies do not necessarily mean that the
calculated gaps are incorrect. Instead, it may be that the pairing gap
is not determined properly from odd-even mass differences, as discussed
in the text.  In particular, when large errors in this figure occur in
the same region where large errors occur in Fig.~\ref{errnln},
sudden shape transitions are probably responsible.
 
\item[Fig.\ \ref{errnln}\hfill]
Discrepancy between experimental neutron pairing gaps determined from
fourth-order odd-even mass differences and microscopic pairing gaps $\Delta_{{\rm
LN}_{\rm n}}= \Delta_{\rm n} +\lambda_{\rm 2n}$ obtained in the Lipkin-Nogami
model.  Magic proton and neutron numbers are indicated by pairs of
thin, parallel lines. Fairly large discrepancies occur in several
places. However, these discrepancies do not necessarily mean that the
calculated gaps are incorrect. Instead, it may be that the pairing gap
is not determined properly from odd-even mass differences, as discussed
in the text.  In particular, when large errors in this figure occur in
the same region where large errors occur in Fig.~\ref{errpln},
sudden shape transitions are probably responsible.
 
\item[Fig.\ \ref{s1nod92a}\hfill]
Neutron separation energy for odd-neutron nuclei. Each odd-neutron
nucleus is represented by a color field {\it one} unit high and {\it
two} units wide.  Black squares denote $\beta$-stable nuclei.  Where
available, experimental masses were used to determine the location of
$\beta$-stable nuclei; otherwise, calculated masses were used.  The
region of known nuclei is enclosed by a jagged black line, and magic proton
and neutron numbers are indicated by pairs of thin, parallel lines. The
region of the nuclear chart where the $r$-process occurs is
approximately 2.0 MeV $<S_{\rm 1n}<$ 3.0 MeV for odd $N$,  which in this figure
appears in light blue. Recent experiments have reached the
$r$-process region at both  $N=50$ and $N=82$.
 
\item[Fig.\ \ref{s1nev92a}\hfill]
Neutron separation energy for even-neutron nuclei. Each even-neutron
nucleus is represented by a color field {\it one} unit high and {\it
two} units wide. Black squares denote $\beta$-stable nuclei.  Where
available, experimental masses were used to determine the location of
$\beta$-stable nuclei; otherwise, calculated masses were used.  The
region of known nuclei is enclosed by a jagged black line, and magic proton and
neutron numbers are indicated by pairs of thin, parallel lines.
 
\item[Fig.\ \ref{s2n92a}\hfill]
Two-neutron separation energy for odd- and even-neutron nuclei. Each
nucleus is represented by a color field one unit high and
one unit wide. Black squares denote $\beta$-stable nuclei.
Where available, experimental masses were used to determine the
location of $\beta$-stable nuclei; otherwise, calculated masses were used.
The region of known nuclei is enclosed by a jagged black line, and
magic proton and neutron numbers are indicated by pairs of
thin, parallel lines.
 
\item[Fig.\ \ref{errs1n92a}\hfill]
Discrepancy between experimental and calculated one-neutron separation
energies.  Magic proton and neutron numbers are indicated by pairs of
thin, parallel lines.  Large discrepancies occur where there are large
changes in the mass error between neighboring nuclei, namely
at magic numbers and in the light region of nuclei. In the
deformed rare-earth and actinide regions the discrepancy is very small.
 
\item[Fig.\ \ref{s1pod92a}\hfill]
Proton separation energy for odd-proton nuclei. Each odd-proton
nucleus is represented by a color field {\it one} unit wide and {\it
two} units high.  Black squares denote $\beta$-stable nuclei.  Where
available, experimental masses were used to determine the location of
$\beta$-stable nuclei; otherwise, calculated masses were used.  The
region of known nuclei, which is enclosed by a jagged black line,
extends in several places   to where proton
emission is energetically allowed from odd-proton nuclei.
Magic proton and    
neutron numbers are indicated by pairs of thin, parallel lines.
 
\item[Fig.\ \ref{s1pev92a}\hfill]
Proton separation energy for even-proton nuclei. Each even-proton
nucleus is represented by a color field {\it one} unit wide and {\it
two} units high.  Black squares denote $\beta$-stable nuclei.  Where
available, experimental masses were used to determine the location of
$\beta$-stable nuclei; otherwise, calculated masses were used.  The
region of known nuclei, which is enclosed by a jagged black line,
nowhere extends  to where proton
emission  is energetically allowed from even-proton nuclei.
Magic proton and    
neutron numbers are indicated by pairs of thin, parallel lines.
 
\item[Fig.\ \ref{s2p92a}\hfill]
Two-proton separation energy for odd- and even-proton nuclei.
Each  nucleus is represented by a color field one unit wide and
one unit high.  Black squares denote $\beta$-stable nuclei.  Where
available, experimental masses were used to determine the location of
$\beta$-stable nuclei; otherwise, calculated masses were used.  The
region of known nuclei, which is enclosed by a jagged black line,
barely extends  to where two-proton
emission  is energetically allowed.
Magic proton and    
neutron numbers are indicated by pairs of thin, parallel lines.
 
\item[Fig.\ \ref{errs1p92a}\hfill]
Discrepancy between experimental and calculated one-proton separation
energies.  Magic proton and neutron numbers are indicated by pairs of
thin, parallel lines.  Large discrepancies occur where there are large
changes in the mass error between neighboring nuclei, namely
at magic numbers and in the light region of nuclei. In the
deformed rare-earth and actinide regions the discrepancy is very small.
 
\item[Fig.\ \ref{qal92a}\hfill]
Energy released in $\alpha$ decay.  Black squares denote $\beta$-stable
nuclei.  Where available, experimental masses were used to determine
the location of $\beta$-stable nuclei; otherwise, calculated masses
were used.  The region of known nuclei is enclosed by a jagged black line, and
magic proton and neutron numbers are indicated by pairs of thin,
parallel lines.  The calculations show that $Q_{\alpha}$ is
in the range 6--9 MeV in the heaviest known region and 9--12 MeV in the
deformed superheavy island surrounding $^{272}$110.
 
\item[Fig.\ \ref{talp92a}\hfill]
Global $\alpha$-decay  half-life calculated from a semi-empirical
relationship between $T_{\alpha}$ and $Q_{\alpha}$.
 Black squares denote $\beta$-stable
nuclei.  Where available, experimental masses were used to determine
the location of $\beta$-stable nuclei; otherwise, calculated masses
were used.  The region of known nuclei is enclosed by a jagged black line, and
magic proton and neutron numbers are indicated by pairs of thin,
parallel lines.     
 
\item[Fig.\ \ref{betlif}\hfill]
Global microscopic $\beta$-decay half-life for allowed
Gamow-Teller transitions.  Black squares denote $\beta$-stable nuclei.
Where available, experimental masses were used to determine the location
of $\beta$-stable nuclei; otherwise, calculated masses were used.  The
region of known nuclei is enclosed by a jagged black line, and magic proton and
neutron numbers are indicated by pairs of thin, parallel lines.  Above
the black squares the calculated combined half-life with respect to
$\beta^+$ decay and electron capture is plotted; below the black
squares the half-life with respect to $\beta^-$ decay is plotted.
 
\item[Fig.\ \ref{rproc}\hfill]
Features of the $r$-process.  Black squares denote $\beta$-stable
nuclei.  Where available, experimental masses were used to determine
the location of $\beta$-stable nuclei; otherwise, calculated masses
were used.  The colored region in the main graph shows calculated
half-life with respect to $\beta^-$ decay.  The jagged black line gives
the right-hand boundary of the region of known nuclei.  The thick
magenta line represents $S_{\rm 1n}=2.4$ MeV\@,  which is the
approximate location of the $r$-process path for a particular set of
stellar conditions.  The magenta squares in the region of
$\beta$-stable nuclei are created in decay from the $r$-process line.
The solar $r$-process abundance shown in the insert is plotted versus
the mass number $A$, whose axis is  curved slightly to follow the line
of $\beta$-stability. A line perpendicular to the valley of
$\beta$-stability and originating at a particular mass value crosses
the $A$ axis of the insert plot at right angles at this value and also
passes through the circle giving the abundance for this $A$ value.

\item[Fig.\ \ref{spincoa92a}\hfill]
Comparison of calculated and experimental nuclear ground-state spins
and parities        
for odd-even nuclei in the light and medium-mass regions.
Spherical assignments are used in the
calculations when $|\epsilon_2|<0.15$. Many of the discrepancies occur
in transition regions between spherical and deformed nuclei or where
several levels are grouped close together.
 
\item[Fig.\ \ref{spincob92a}\hfill]
Similar to Fig.~\ref{spincoa92a} but for the heavy region.  The
discrepancies in the heaviest part of the actinide  region occur
because here several neutron single-particle levels are grouped very
close together.     
 
\item[Fig.\ \ref{masd92a}\hfill]
Comparison of experimental  and calculated microscopic corrections for
1654 nuclei in the 1992 version of the finite-range
droplet model.$\,^{1,46})$\ess
The bottom part showing the difference between these two
quantities is equivalent to the difference between measured and
calculated ground-state masses. There are almost no systematic errors
remaining for nuclei with $N \geq 65$, for which region the
theoretical error is only
0.448  MeV\@.       
 
\item[Fig.\ \ref{bsrb95}\hfill]
Calculated Gamow-Teller $\beta$-strength function for $^{95}$Rb.
Whereas for the calculation of the quantities in the Table  we use the
shapes obtained in our mass calculation,$\,^{1})$\mss in this
figure we use the spherical shape because we wish to show a typical
spherical $\beta$-strength function and because experimentally
$^{95}$Rb is spherical.  The arrows with wide heads denote successive
neutron-separation energies in the daughter, and the arrow with a thin
head denotes the value of $Q_{\beta}$.  The calculated strength
containing only a few large peaks in the low-energy region is typical
of a spherical nucleus.  Since all the strength is calculated to be
above the one-neutron separation energy, the theoretical probability
$P_{\rm n}$ for $\beta$-delayed neutron emission is 100\%.
Experimentally, there is a large peak in the GT strength function in
the region 3.5--4.0 MeV\@,  so experimentally the $\beta$-delayed
neutron emission probability is only 8.5\%.
 
\item[Fig.\ \ref{bsrb99}\hfill]
Calculated Gamow-Teller $\beta$-strength function for $^{99}$Rb.
Here we use the shape obtained in our mass calculation.
The strength function is typical of that of a deformed nucleus.
Because there is a higher likelihood of significant strength in
the low-energy region for deformed nuclei than for spherical
nuclei there is a characteristic, large decrease in the
$\beta$-decay half-life at the shape transition.
Since there is considerable
strength below the neutron-emission threshold the $\beta$-delayed
neutron-emission probability is low.

\item[Fig.\ \ref{betlifmn}\hfill]
Ratios between calculated and experimental half-lives for
$\beta^-$  decay as functions
of neutron number $N$. There are no systematic effects
versus $N$.         
 
\item[Fig.\ \ref{betlifmt}\hfill]
Ratios between calculated and experimental half-lives for
$\beta^-$ decay as functions
of the experimental half-life for $\beta^-$  decay.
As expected, we find a very strong correlation between the error and
the experimental $\beta$-decay half-life. The  correlation
is such that we can expect fairly reliable half-life calculations
far from $\beta$-stability, in the region of interest for
astrophysical $r$-process calculations.
An analysis of the results in this figure is presented in Table
\ref{tabbetm} and is also discussed in the text.
 
\item[Fig.\ \ref{betlifmq}\hfill]
Ratios between calculated and experimental half-lives for
$\beta^-$ decay as functions
of $Q_{\rm \beta}$. The discrepancy
is expected to be larger for low values of $Q_{\rm \beta}$
because the calculated half-life is most sensitive here to
errors in the positions of the peaks in the strength functions.
 
\item[Fig.\ \ref{betlifpn}\hfill]
Ratios between calculated and experimental half-lives for
$\beta^+$ decay and electron capture  as functions
of neutron number $N$. There are no systematic effects
versus $N$.

\item[Fig.\ \ref{betlifpt}\hfill]
Ratios between calculated and experimental half-lives for
$\beta^+$ decay and electron capture  as functions
of the experimental half-life for $\beta^+$ decay and electron capture.
Surprisingly, there is no strong correlation between the error and
$T_{\beta,{\rm exp}}$,
except for odd-odd nuclei.
An analysis of the results in this figure is presented in Table
\ref{tabbetp} and is also discussed in the text.
 
\item[Fig.\ \ref{betlifpq}\hfill]
Ratios between calculated and experimental half-lives for
$\beta^+$ decay and electron capture  as functions
of $Q_{\beta}$. Only the error for odd-odd nuclei
is correlated with $Q_{\beta}$.
 
\item[Fig.\ \ref{qcomp1}\hfill]
Comparison between experimental and calculated energy releases $Q_{\alpha}$ for the
$N=154$ and 155 isotonic chains. The experimental data are from Fig.~6
of M\"{u}nzenberg {\it et al.\/},$\,^{42})$\ess
where the data are compared to
predictions of the FRLDM (1988). The new results presented here agree
with the data much better. The improvement is due partly to the inclusion
of the $\epsilon_6$ shape degree of freedom.
 
\item[Fig.\ \ref{devmfrdm}\hfill]
Calculation to show the reliability of the FRDM (1992) in new regions of nuclei.
The FRDM (1992) was adjusted to 1654 masses known in
1989.$\,^{2})$\ess The figure shows the deviations between experimental
and calculated masses for 217 new nuclei whose
masses were measured between 1989
and 1993.$\,^{3,4})$\ess  The error is
4\% {\it smaller} in the new region compared to that in the region
where the model constants were adjusted. There are no
systematic effects visible in the figure.
 
\item[Fig.\ \ref{devmetf}\hfill]
Calculation to show the reliability of the ETFSI-1 (1992) in new regions of
nuclei.  The ETFSI-1 (1992) was adjusted to  masses known in
1988.$\,^{49})$\ess The figure shows the deviations between experimental
and calculated masses for 210 new nuclei whose
masses were measured between
1989 and 1993.$\,^{3,4})$\ess  The error
is 10\% larger in the new region compared to that for the
1989 data set$\,^{2})$ we have available, which is only
marginally different from the 1988 data set to which the model constants
were adjusted. There are no systematic effects visible in
the figure.         
 
\item[Fig.\ \ref{devmgro}\hfill]
Similar to Figs.~\ref{devmfrdm} and \ref{devmetf}
but for the mass formula of von Groote
{\it et al.\/}$\,^{50})$\ess With its
postulated shell corrections and more adjustable constants than in
our model with {\it calculated\/} shell corrections, the error in the
new region is 104\% larger than for a 1977 set of measured
masses,$\,^{51})$\mss which is only marginally different from
the set of masses where the  constants were adjusted. There is
also a systematic increase in the error with increasing distance from $\beta$
stability.

\item[Fig.\ \ref{devs1nfrdm}\hfill]
Calculation to show the reliability of one-neutron separation energies
obtained from the FRDM (1992) in new regions of nuclei.
There are no systematic effects visible in the figure.
 
\item[Fig.\ \ref{devs1netf}\hfill]
Calculation to show the reliability of one-neutron separation energies
obtained from the ETFSI-1 (1992) model in new regions of nuclei.
There are no systematic effects visible in the figure,
apart from an odd-even staggering related to problems
in the pairing part of the ETFSI-1 (1992) model.

\item[Fig.\ \ref{devs1ngro}\hfill]
Calculation to show the reliability of one-neutron separation energies
obtained from the mass formula of von Groote {\it et al\/}.\ in new regions of nuclei.
On the neutron-rich side the calculated one-neutron separation energies
are systematically too high by 0.29 MeV\@.

\item[Fig.\ \ref{devqbmfrdm}\hfill]
Calculation to show the reliability of energy releases
for $\beta^-$ decay 
obtained from the FRDM (1992) in new regions of nuclei.
There are no systematic effects visible in the figure.
 
\item[Fig.\ \ref{devqbmetf}\hfill]
Calculation to show the reliability of energy releases
for $\beta^-$ decay 
obtained from the ETFSI-1 (1992) model in new regions of nuclei.

\item[Fig.\ \ref{devqbmgro}\hfill]
Calculation to show the reliability of energy releases for $\beta^-$ decay
obtained from the mass formula of von Groote {\it et al\/}.\ in
new regions of nuclei. On the neutron-rich side the
calculated energy releases are systematically too low by 0.52 MeV\@.
Corresponding $\beta$-decay rates
based on these values would be too slow.
 
\item[Fig.\ \ref{devmtf}\hfill]
Calculation to show the reliability of the TF (1994) model in new regions of
nuclei.  The TF (1994) model
was adjusted to 1654 masses known in
1989.$\,^{2})$\ess The figure shows the deviations between experimental
and calculated masses for 217 new nuclei whose
masses were measured between 1989
and 1993.$\,^{3,4})$\ess  The error is
15\% larger in the new region compared to that in the region
where the model constants were adjusted. There are no
systematic effects visible in the figure.
 
\item[Fig.\ \ref{devmsyst}\hfill]
Calculation to show the reliability of the
Wapstra-Audi systematic-trend mass model in new regions of
nuclei.  The systematic-trend masses were provided
 in the 1989 midstream mass evaluation.$\,^{2})$\ess
The figure shows the deviations between experimental
and systematic masses for 187 new nuclei that were
given by systematic trends in the 1989 evaluation and whose
masses were measured between 1989
and 1993.$\,^{3,4})$\ess   There are no
systematic effects visible in the figure.

\item[Fig.\ \ref{comp3mod}\hfill]
Comparison between energy releases  $Q_{\alpha}$ obtained in the
FRDM (1992), ETFSI-1 (1992) model, and FDSM (1992)  and recent
experimental data   
for the heaviest known element.$\,^{30})$\ess
When several  values of $Q_{\alpha}$ were measured we choose for
the figure the highest value.

\item[Fig.\ \ref{rpprocfig}\hfill]
The hot hydrogen burning cycles,$\,^{57})$\mss
typically consisting of three
proton captures, two $\beta^+$ decays, and a closing
(p,$\alpha$) reaction. Break-out towards heavier nuclei occurs only via
the (p,$\gamma$)/(p,$\alpha$)
branching at the cycle closings. Because of the $^{18}$F(p,$\alpha$) reaction
no OFNe cycle exists, which would otherwise connect the CNO and NeNaMg cycles.
 
\item[Fig.\ \ref{rabu1}\hfill]
Calculated $r$-process abundances (solid lines) compared
to measured values (solid circles). For both the upper and lower
parts of the figure 
$\beta$-decay half-lives and delayed-neutron emission
probabilities are calculated in a QRPA model based on
folded-Yukawa single-particle energies, but experimental
information has been used when available. In the upper part of the
figure the $r$-process path was determined from the
FRDM (1991),$\,^{45,67})$\mss
and in the lower part of the figure it was determined
from the            
preliminary, privately circulated version
of  the ETFSI-1 (1992) model.$\,^{47,48})$
 
\item[Fig.\ \ref{rabu2}\hfill]
Static steady-flow calculations of the $r$-process
abundance $N_{r,\odot}$ for the
$135\leq A\leq 195$ mass region, the so-called ``third-component.''
The right-hand part of the figure shows the best fit obtained with $T_{\beta}$ and
$P_{\rm n}$ values from our QRPA model. The upper left-hand part
of the figure shows $N_{r,\odot}$  for the same
$T_9$-$n_{\rm n}$ conditions obtained with the on-the-average five-times
longer $T_{\beta}$ values from the gross theory.$\,^{73})$\ess
As can be seen from the lower left-hand part of the figure, with these $T_{\beta}$ values
reasonable fits for the actinide region require neutron densities of
$10^{25}$~cm$^{-3}$, which are difficult to obtain in the
hot-entropy-bubble $r$-process scenario.
 
\item[Fig. \ref{p008016}\hfill]
Calculated proton single-particle level diagram for
nuclei in the vicinity of $^{16}_{\phantom{0}8}$O$_{8}^{\phantom{0}}$.
 
\item[Fig. \ref{n008016}\hfill]
Calculated neutron single-particle level diagram for
nuclei in the vicinity of $^{16}_{\phantom{0}8}$O$_{8}^{\phantom{0}}$.
 
\item[Fig. \ref{p020034}\hfill]
Calculated proton single-particle level diagram for proton-rich
nuclei in the vicinity of $^{34}_{20}$Ca$_{14}^{\phantom{00}}$.
 
\item[Fig. \ref{n020034}\hfill]
Calculated neutron single-particle level diagram for proton-rich
nuclei in the vicinity of $^{34}_{20}$Ca$_{14}^{\phantom{00}}$.
 
\item[Fig. \ref{p020044}\hfill]
Calculated proton single-particle level diagram for
nuclei in the vicinity of $^{44}_{20}$Ca$_{24}^{\phantom{00}}$.
 
\item[Fig. \ref{n020044}\hfill]
Calculated neutron single-particle level diagram for
nuclei in the vicinity of $^{44}_{20}$Ca$_{24}^{\phantom{00}}$.
 
\item[Fig. \ref{p020054}\hfill]
Calculated proton single-particle level diagram for neutron-rich
nuclei in the vicinity of $^{54}_{20}$Ca$_{34}^{\phantom{00}}$.
 
\item[Fig. \ref{n020054}\hfill]
Calculated neutron single-particle level diagram for neutron-rich
nuclei in the vicinity of $^{54}_{20}$Ca$_{34}^{\phantom{00}}$.
 
\item[Fig. \ref{p028050}\hfill]
Calculated proton single-particle level diagram for proton-rich
nuclei in the vicinity of $^{50}_{28}$Ni$_{22}^{\phantom{00}}$.
 
\item[Fig. \ref{n028050}\hfill]
Calculated neutron single-particle level diagram for proton-rich
nuclei in the vicinity of $^{50}_{28}$Ni$_{22}^{\phantom{00}}$.
 
\item[Fig. \ref{p028064}\hfill]
Calculated proton single-particle level diagram for
nuclei in the vicinity of $^{64}_{28}$Ni$_{36}^{\phantom{00}}$.
 
\item[Fig. \ref{n028064}\hfill]
Calculated neutron single-particle level diagram for
nuclei in the vicinity of $^{64}_{28}$Ni$_{36}^{\phantom{00}}$.
 
\item[Fig. \ref{p028078}\hfill]
Calculated proton single-particle level diagram for neutron-rich
nuclei in the vicinity of $^{78}_{28}$Ni$_{50}^{\phantom{00}}$.
 
\item[Fig. \ref{n028078}\hfill]
Calculated neutron single-particle level diagram for neutron-rich
nuclei in the vicinity of $^{78}_{28}$Ni$_{50}^{\phantom{00}}$.
 
\item[Fig. \ref{p040080}\hfill]
Calculated proton single-particle level diagram for proton-rich
nuclei in the vicinity of $^{80}_{40}$Zr$_{40}^{\phantom{00}}$.
The hexacontratetrapole deformation $\epsilon_6$ is zero for oblate
shapes and is given by $\epsilon_6=-0.1\epsilon_2$ for prolate shapes.

\item[Fig. \ref{n040080}\hfill]
Calculated neutron single-particle level diagram for proton-rich
nuclei in the vicinity of $^{80}_{40}$Zr$_{40}^{\phantom{00}}$.
The hexacontratetrapole deformation $\epsilon_6$ is zero for oblate
shapes and is given by $\epsilon_6=-0.1\epsilon_2$ for prolate shapes.
 
\item[Fig. \ref{p040090}\hfill]
Calculated proton single-particle level diagram for
nuclei in the vicinity of $^{90}_{40}$Zr$_{50}^{\phantom{00}}$.
 
\item[Fig. \ref{n040090}\hfill]
Calculated neutron single-particle level diagram for
nuclei in the vicinity of $^{90}_{40}$Zr$_{50}^{\phantom{00}}$.
 
\item[Fig. \ref{p040106}\hfill]
Calculated proton single-particle level diagram for neutron-rich
nuclei in the vicinity of $^{106}_{\phantom{0}40}$Zr$_{66}^{\phantom{00}}$.
The hexacontratetrapole deformation $\epsilon_6$ is zero for oblate
shapes and is given by $\epsilon_6=-0.1\epsilon_2$ for prolate shapes.
 
\item[Fig. \ref{n040106}\hfill]
Calculated neutron single-particle level diagram for neutron-rich
nuclei in the vicinity of $^{106}_{\phantom{0}40}$Zr$_{66}^{\phantom{00}}$.
The hexacontratetrapole deformation $\epsilon_6$ is zero for oblate
shapes and is given by $\epsilon_6=-0.1\epsilon_2$ for prolate shapes.
 
\item[Fig. \ref{p050100}\hfill]
Calculated proton single-particle level diagram for proton-rich
nuclei in the vicinity of $^{100}_{\phantom{0}50}$Sn$_{50}^{\phantom{00}}$.
 
\item[Fig. \ref{n050100}\hfill]
Calculated neutron single-particle level diagram for proton-rich
nuclei in the vicinity of $^{100}_{\phantom{0}50}$Sn$_{50}^{\phantom{00}}$.
 
\item[Fig. \ref{p050116}\hfill]
Calculated proton single-particle level diagram for
nuclei in the vicinity of $^{116}_{\phantom{0}50}$Sn$_{66}^{\phantom{00}}$.
 
\item[Fig. \ref{n050116}\hfill]
Calculated neutron single-particle level diagram for
nuclei in the vicinity of $^{116}_{\phantom{0}50}$Sn$_{66}^{\phantom{00}}$.
 
\item[Fig. \ref{p050132}\hfill]
Calculated proton single-particle level diagram for neutron-rich
nuclei in the vicinity of $^{132}_{\phantom{0}50}$Sn$_{82}^{\phantom{00}}$.
 
\item[Fig. \ref{n050132}\hfill]
Calculated neutron single-particle level diagram for neutron-rich
nuclei in the vicinity of $^{132}_{\phantom{0}50}$Sn$_{82}^{\phantom{00}}$.
 
\item[Fig. \ref{p062132}\hfill]
Calculated proton single-particle level diagram for proton-rich
nuclei in the vicinity of $^{132}_{\phantom{0}62}$Sm$_{70}^{\phantom{00}}$.
 
\item[Fig. \ref{n062132}\hfill]
Calculated neutron single-particle level diagram for proton-rich
nuclei in the vicinity of $^{132}_{\phantom{0}62}$Sm$_{70}^{\phantom{00}}$.
 
\item[Fig. \ref{p062158}\hfill]
Calculated proton single-particle level diagram for neutron-rich
nuclei in the vicinity of $^{158}_{\phantom{0}62}$Sm$_{96}^{\phantom{00}}$.
 
\item[Fig. \ref{n062158}\hfill]
Calculated neutron single-particle level diagram for neutron-rich
nuclei in the vicinity of $^{158}_{\phantom{0}62}$Sm$_{96}^{\phantom{00}}$.
 
\item[Fig. \ref{p074156}\hfill]
Calculated proton single-particle level diagram for proton-rich
nuclei in the vicinity of $^{156}_{\phantom{0}74}$W$_{82}^{\phantom{00}}$.
 
\item[Fig. \ref{n074156}\hfill]
Calculated neutron single-particle level diagram for proton-rich
nuclei in the vicinity of $^{156}_{\phantom{0}74}$W$_{82}^{\phantom{00}}$.
 
\item[Fig. \ref{p074190}\hfill]
Calculated proton single-particle level diagram for neutron-rich
nuclei in the vicinity of $^{190}_{\phantom{0}74}$W$_{116}^{\phantom{00}}$.
 
\item[Fig. \ref{n074190}\hfill]
Calculated neutron single-particle level diagram for neutron-rich
nuclei in the vicinity of $^{190}_{\phantom{0}74}$W$_{116}^{\phantom{00}}$.
 
\item[Fig. \ref{p082180}\hfill]
Calculated proton single-particle level diagram for proton-rich
nuclei in the vicinity of $^{180}_{\phantom{0}82}$Pb$_{98}^{\phantom{00}}$.
 
\item[Fig. \ref{n082180}\hfill]
Calculated neutron single-particle level diagram for proton-rich
nuclei in the vicinity of $^{180}_{\phantom{0}82}$Pb$_{98}^{\phantom{00}}$.
 
\item[Fig. \ref{p082208}\hfill]
Calculated proton single-particle level diagram for
nuclei in the vicinity of $^{208}_{\phantom{0}82}$Pb$_{126}^{\phantom{00}}$.
 
\item[Fig. \ref{n082208}\hfill]
Calculated neutron single-particle level diagram for
nuclei in the vicinity of $^{208}_{\phantom{0}82}$Pb$_{126}^{\phantom{00}}$.
 
\item[Fig. \ref{p082238}\hfill]
Calculated proton single-particle level diagram for neutron-rich
nuclei in the vicinity of $^{238}_{\phantom{0}82}$Pb$_{156}^{\phantom{00}}$,
which is located on the $r$-process path
and is calculated to be deformed
in its ground state.
 
\item[Fig. \ref{n082238}\hfill]
Calculated neutron single-particle level diagram for neutron-rich
nuclei in the vicinity of $^{238}_{\phantom{0}82}$Pb$_{156}^{\phantom{00}}$,
which is located on the $r$-process path
and is calculated to be deformed
in its ground state.
 
\item[Fig. \ref{p090216}\hfill]
Calculated proton single-particle level diagram for proton-rich
nuclei in the vicinity of $^{216}_{\phantom{0}90}$Th$_{126}^{\phantom{00}}$.
 
\item[Fig. \ref{n090216}\hfill]
Calculated neutron single-particle level diagram for proton-rich
nuclei in the vicinity of $^{216}_{\phantom{0}90}$Th$_{126}^{\phantom{00}}$.
 
\item[Fig. \ref{p090232}\hfill]
Calculated proton single-particle level diagram for
nuclei in the vicinity of $^{232}_{\phantom{0}90}$Th$_{142}^{\phantom{00}}$.
The hexacontratetrapole deformation $\epsilon_6$ is zero for oblate
shapes and is given by $\epsilon_6=0.1\epsilon_2$ for prolate shapes.
 
\item[Fig. \ref{n090232}\hfill]
Calculated neutron single-particle level diagram for
nuclei in the vicinity of $^{232}_{\phantom{0}90}$Th$_{142}^{\phantom{00}}$.
The hexacontratetrapole deformation $\epsilon_6$ is zero for oblate
shapes and is given by $\epsilon_6=0.1\epsilon_2$ for prolate shapes.
 
\item[Fig. \ref{p100240}\hfill]
Calculated proton single-particle level diagram for proton-rich
nuclei in the vicinity of $^{240}_{100}$Fm$_{140}^{\phantom{00}}$.
The hexacontratetrapole deformation $\epsilon_6$ is zero for oblate
shapes and is given by $\epsilon_6=0.15\epsilon_2$ for prolate shapes.
 
\item[Fig. \ref{n100240}\hfill]
Calculated neutron single-particle level diagram for proton-rich
nuclei in the vicinity of $^{240}_{100}$Fm$_{140}^{\phantom{00}}$.
The hexacontratetrapole deformation $\epsilon_6$ is zero for oblate
shapes and is given by $\epsilon_6=0.15\epsilon_2$ for prolate shapes.
 
\item[Fig. \ref{p100252}\hfill]
Calculated proton single-particle level diagram for
nuclei in the vicinity of $^{252}_{100}$Fm$_{152}^{\phantom{00}}$.
The hexacontratetrapole deformation $\epsilon_6$ is zero for oblate
shapes and is given by $\epsilon_6=0.2\epsilon_2$ for prolate shapes.

\item[Fig. \ref{n100252}\hfill]
Calculated neutron single-particle level diagram for
nuclei in the vicinity of $^{252}_{100}$Fm$_{152}^{\phantom{00}}$.
The hexacontratetrapole deformation $\epsilon_6$ is zero for oblate
shapes and is given by $\epsilon_6=0.2\epsilon_2$ for prolate shapes.
 
\item[Fig. \ref{p100262}\hfill]
Calculated proton single-particle level diagram for
nuclei in the vicinity of $^{262}_{100}$Fm$_{162}^{\phantom{00}}$.
 
\item[Fig. \ref{n100262}\hfill]
Calculated neutron single-particle level diagram for
nuclei in the vicinity of $^{262}_{100}$Fm$_{162}^{\phantom{00}}$.
 
\item[Fig. \ref{p110272}\hfill]
Calculated proton single-particle level diagram for  superheavy
nuclei in the vicinity of $^{272}$110$_{162}^{\phantom{00}}$,
which is calculated to be deformed in its ground state.
 
\item[Fig. \ref{n110272}\hfill]
Calculated neutron single-particle level diagram for superheavy
nuclei in the vicinity of $^{272}$110$_{162}^{\phantom{00}}$,
which is calculated to be deformed in its ground state.
 
\item[Fig. \ref{p114298}\hfill]
Calculated proton single-particle level diagram for superheavy
nuclei in the vicinity of $^{298}$114$_{184}^{\phantom{00}}$,
which is calculated to be spherical in its ground state.
 
\item[Fig. \ref{n114298}\hfill]
Calculated neutron single-particle level diagram for superheavy
nuclei in the vicinity of $^{298}$114$_{184}^{\phantom{00}}$,
which is calculated to be spherical in its ground state.
 
\item[Fig. \ref{p124308}\hfill]
Calculated proton single-particle level diagram for superheavy
nuclei in the vicinity of $^{308}$124$_{184}^{\phantom{00}}$,
which is calculated to be spherical in its ground state.
 
\item[Fig. \ref{n124308}\hfill]
Calculated neutron single-particle level diagram for  superheavy
nuclei in the vicinity of $^{308}$124$_{184}^{\phantom{00}}$,
which is calculated to be spherical in its ground state.

\end{list}          
\newpage            
\setcounter{page}{141}
\begin{center}      
{\bf EXPLANATION OF TABLE}
\end{center}        
{\bf Table. Calculated Nuclear Ground-State Properties}\\
\begin{list}        
{\Roman{bean}}{\usecounter{bean}
\setlength{\leftmargin}{1.0in}
\setlength{\rightmargin}{0.0in}
\setlength{\labelwidth}{0.75in}
\setlength{\labelsep}{0.25in}
\setlength{\itemsep}{0.01in}
}                   
\item[{\boldmath $Z$} \hfill ]
Proton number. The Table is ordered by
increasing          
proton number. The corresponding chemical symbol
of each named element is given in parentheses.
For consistency we use the same naming scheme as in our earlier
publication,$\,^{1})$\mss although the names of some
of the heavier elements may be officially changed in the future.
 
\item[$N$ \hfill ]  
Neutron number.     
 
\item[$A$ \hfill ]  
Mass number.

\item[$\Omega_{\rm p}^{\pi}$  \hfill ]
Projection of the odd-proton angular momentum along the
symmetry axis and parity of the wave function.
 
\item[$\Omega_{\rm n}^{\pi}$   \hfill ]
Projection of the odd-neutron angular momentum along the
symmetry axis and parity of the wave function.
 
\item[$\Delta_{{\rm LN}_{\rm p}}$  \hfill ]
Pairing gap for protons in the Lipkin-Nogami model,
given by $\Delta_{\rm p}+ \lambda_{2{\rm p}}$.
 
\item[$\Delta_{{\rm LN}_{\rm n}}$   \hfill ]
Pairing gap for neutrons in the Lipkin-Nogami model,
given by $\Delta_{\rm n}+ \lambda_{2{\rm n}}$.
 
\item[$E_{\rm bind}$   \hfill ]
Total binding energy.
 
\item[$S_{\rm 1n}$  \hfill ]
One-neutron separation energy.
 
\item[$S_{\rm 2n}$   \hfill ]
Two-neutron separation energy.
 
\item[$P_{A}$   \hfill ]
Probability for producing a final nucleus with mass number $A$
following $\beta$ decay and delayed neutron emission.
 
\item[$P_{A-1}$  \hfill ]
Probability for producing a final nucleus with mass number $A-1$
following $\beta$ decay and delayed neutron emission.
 
\item[$P_{A-2}$   \hfill ]
Probability for producing a final nucleus with mass number $A-2$
following $\beta$ decay and delayed neutron emission.
 
\item[$Q_{\beta}$   \hfill ]
Energy released in $\beta$ decay.
 
\item[$T_{\beta}$  \hfill ]
Half-life  with respect to Gamow-Teller
$\beta$ decay.      
 
\item[$S_{\rm 1p}$  \hfill ]
One-proton separation energy.
 
\item[$S_{\rm 2p}$  \hfill ]
Two-proton separation energy.
 
\item[$Q_{\rm \alpha}$  \hfill ]
Energy released in $\alpha$ decay.
 
\item[$T_{\alpha}$  \hfill ]
Half-life with respect to $\alpha$ decay.
                                                       \hfill\\[1ex]
\end{list}          
\mbox{ }\\          
\newpage            
\begin{enumerate}   
\item               
    \label{lippdpl2}
\mbox{}             
\item               
    \label{lipndpl2}
\mbox{}             
\item               
    \label{errpln}  
\mbox{}             
\item               
    \label{errnln}  
\mbox{}             
\item               
    \label{s1nod92a}
\mbox{}             
\item               
    \label{s1nev92a}
\mbox{}             
\item               
    \label{s2n92a}  
\mbox{}             
\item               
    \label{errs1n92a}
\mbox{}             
\item               
    \label{s1pod92a}
\mbox{}             
\item               
    \label{s1pev92a}
\mbox{}             
\item               
    \label{s2p92a}  
\mbox{}             
\item               
    \label{errs1p92a}
\mbox{}             
\item               
    \label{qal92a}  
\mbox{}             
\item               
    \label{talp92a} 
\mbox{ }            
\item               
    \label{betlif}  
\mbox{}             
\item               
    \label{rproc}   
\mbox{}             
\item               
    \label{spincoa92a}
\mbox{ }            
\item               
    \label{spincob92a}
\mbox{ }            
\item{ }            
  \label{masd92a}   
\mbox{ }            
\item               
    \label{bsrb95}  
\mbox{ }            
\item               
    \label{bsrb99}  
\mbox{ }            
\item               
    \label{betlifmn}
\mbox{ }            
\item               
    \label{betlifmt}
\mbox{ }            
\item               
    \label{betlifmq}
\mbox{ }            
\item               
    \label{betlifpn}
\mbox{ }            
\item               
    \label{betlifpt}
\mbox{ }            
\item               
    \label{betlifpq}
\mbox{ }            
\item               
    \label{qcomp1}  
\mbox{ }            
\item               
    \label{devmfrdm}
\mbox{ }            
\item               
    \label{devmetf} 
\mbox{ }            
\item               
    \label{devmgro} 
\mbox{ }            
\item               
    \label{devs1nfrdm}
\mbox{ }            
\item               
    \label{devs1netf}
\mbox{ }            
\item               
    \label{devs1ngro}
\mbox{ }            
\item               
    \label{devqbmfrdm}
\mbox{ }            
\item               
    \label{devqbmetf}
\mbox{ }            
\item               
    \label{devqbmgro}
\mbox{ }            
\item               
    \label{devmtf}  
\mbox{ }            
\item               
    \label{devmsyst}
\mbox{ }            
\item               
    \label{comp3mod}
\mbox{ }            
\item               
    \label{rpprocfig}
\mbox{ }            
\item               
    \label{rabu1}   
\mbox{ }            
\item               
    \label{rabu2}   
\mbox{ }            
\item               
   \label{p008016}  
\mbox{ }            
\item               
   \label{n008016}  
\mbox{ }            
\item               
   \label{p020034}  
\mbox{ }            
\item               
   \label{n020034}  
\mbox{ }            
\item               
   \label{p020044}  
\mbox{ }            
\item               
   \label{n020044}  
\mbox{ }            
\item               
   \label{p020054}  
\mbox{ }            
\item               
   \label{n020054}  
\mbox{ }            
\item               
   \label{p028050}  
\mbox{ }            
\item               
   \label{n028050}  
\mbox{ }            
\item               
   \label{p028064}  
\mbox{ }            
\item               
   \label{n028064}  
\mbox{ }            
\item               
   \label{p028078}  
\mbox{ }            
\item               
   \label{n028078}  
\mbox{ }            
\item               
   \label{p040080}  
\mbox{ }            
\item               
   \label{n040080}  
\mbox{ }            
\item               
   \label{p040090}  
\mbox{ }            
\item               
   \label{n040090}  
\mbox{ }            
\item               
   \label{p040106}  
\mbox{ }            
\item               
   \label{n040106}  
\mbox{ }            
\item               
   \label{p050100}  
\mbox{ }            
\item               
   \label{n050100}  
\mbox{ }            
\item               
   \label{p050116}  
\mbox{ }            
\item               
   \label{n050116}  
\mbox{ }            
\item               
   \label{p050132}  
\mbox{ }            
\item               
   \label{n050132}  
\mbox{ }            
\item               
   \label{p062132}  
\mbox{ }            
\item               
   \label{n062132}  
\mbox{ }            
\item               
   \label{p062158}  
\mbox{ }            
\item               
   \label{n062158}  
\mbox{ }            
\item               
   \label{p074156}  
\mbox{ }            
\item               
   \label{n074156}  
\mbox{ }            
\item               
   \label{p074190}  
\mbox{ }            
\item               
   \label{n074190}  
\mbox{ }            
\item               
   \label{p082180}  
\mbox{ }            
\item               
   \label{n082180}  
\mbox{ }            
\item               
   \label{p082208}  
\mbox{ }            
\item               
   \label{n082208}  
\mbox{ }            
\item               
   \label{p082238}  
\mbox{ }            
\item               
   \label{n082238}  
\mbox{ }            
\item               
   \label{p090216}  
\mbox{ }            
\item               
   \label{n090216}  
\mbox{ }            
\item               
   \label{p090232}  
\mbox{ }            
\item               
   \label{n090232}  
\mbox{ }            
\item               
   \label{p100240}  
\mbox{ }            
\item               
   \label{n100240}  
\mbox{ }            
\item               
   \label{p100252}  
\mbox{ }            
\item               
   \label{n100252}  
\mbox{ }            
\item               
   \label{p100262}  
\mbox{ }            
\item               
   \label{n100262}  
\mbox{ }            
\item               
   \label{p110272}  
\mbox{ }            
\item               
   \label{n110272}  
\mbox{ }            
\item               
   \label{p114298}  
\mbox{ }            
\item               
   \label{n114298}  
\mbox{ }            
\item               
   \label{p124308}  
\mbox{ }            
\item               
   \label{n124308}  
\mbox{ }            
\end{enumerate}     
\end{document}